\documentclass[aps,prd,amsmath,amssymb,preprintnumbers,nofootinbib,preprint]{revtex4-1}
\pdfoutput=1
\usepackage[utf8]{inputenc} %utf8
\usepackage{graphicx}
\graphicspath{{./figures/}}
\usepackage{url}
\usepackage[bookmarks, pagebackref=false]{hyperref}
\usepackage[usenames,dvipsnames]{xcolor}
\definecolor{orange}{cmyk}{0,0.5,1,0}
\definecolor{rossoCP3}{cmyk}{0,.88,.77,.40}
\definecolor{graa}{rgb}{0.8,0.8,0.8}
\definecolor{blaa}{rgb}{0.2,0.2,0.6}
		\hypersetup{
			colorlinks, 
			bookmarksopen, 
			bookmarksnumbered,
			citecolor=blaa, 		%color of links to bibliography
			linkcolor=rossoCP3,	%color of internal links
			urlcolor=rossoCP3,			%color of external links 
			}
\usepackage{amsthm}
\usepackage{bm}% bold math
\usepackage{bbm}
\usepackage{pxfonts}

\usepackage{amsmath,amssymb,amsfonts}
\usepackage{color}
\usepackage{float}
\usepackage{hyperref}
\usepackage[Symbolsmallscale]{upgreek}
\usepackage{amsmath}
\usepackage{amsfonts}
\usepackage{amssymb,dsfont}
\usepackage{graphicx}
\usepackage{amssymb}
\usepackage[vcentermath]{youngtab}
\usepackage[all]{xy}
\usepackage{pstricks}
\usepackage{dsfont}%
\usepackage{placeins}
\usepackage{xspace}
\usepackage{cancel} % to make the barred text notation

\usepackage{slashed}
\usepackage[caption=false, labelformat=simple, listofformat=subsimple, labelfont=default, margin=5pt, justification=raggedright]{subfig}

	\makeatletter
		\renewcommand{\p@subfigure}{}
	\makeatother

\usepackage{natbib}

\usepackage{feynmf}

\newcommand{\ea}[1]{
\begin{align}
#1
\end{align}
}

\newcommand{\phdagg}{\phantom{\dagger}}

\newcommand{\beq}{\begin{eqnarray}}
\newcommand{\eeq}{\end{eqnarray}}

\newcommand{\bmp}{\noindent\begin{minipage}{16cm}}
\newcommand{\emp}{\end{minipage}\vskip 7mm} % 7mm untightened

\newcommand{\yf}{\tiny \yng(1)}

\DeclareMathOperator{\tr}{Tr}

\def\lsim{\mathrel{\rlap{\lower4pt\hbox{\hskip1pt$\sim$}}
    \raise1pt\hbox{$<$}}}                % less than or approx. symbol
\def\gsim{\mathrel{\rlap{\lower4pt\hbox{\hskip1pt$\sim$}}
    \raise1pt\hbox{$>$}}}                % greater than or approx. symbol

\begin{document}
%%%%%%%%%%%%%%%%%%%%%%%%%%%%%%%%%%%%%%%%%%%%%%%%%%%%%%%%%%%%%%%%%%%%%%%%%%%

\title{\texorpdfstring{\Large\color{rossoCP3} Higgs Critical Exponents   \\  and  \\ Conformal Bootstrap in Four Dimensions}{Higgs Critical Exponents and Conformal Bootstrap in Four Dimensions}}
\author{Oleg {\sc Antipin}$^{\color{rossoCP3}{\clubsuit}}$}
\email{antipin@fi.infn.it} 
\author{Esben {\sc M\o lgaard}$^{\color{rossoCP3}{\diamondsuit}}$}
\email{molgaard@cp3.dias.sdu.dk} 
\author{Francesco {\sc Sannino}$^{\color{rossoCP3}{\diamondsuit}}$}
\email{sannino@cp3.dias.sdu.dk}
\affiliation{{$^{\color{rossoCP3}{\clubsuit}}$ INFN, Sezione di Firenze, Via G. Sansone, 1; I-50019 Sesto Fiorentino, Italy}\\{ $^{\color{rossoCP3}{\diamondsuit}}$\color{rossoCP3} {CP}$^{ \bf 3}${-Origins}} \& the Danish Institute for Advanced Study {\color{rossoCP3}\rm{Danish IAS}},  University of Southern Denmark, Campusvej 55, DK-5230 Odense M, Denmark.}

\begin{abstract}
We investigate relevant properties of composite operators emerging in nonsupersymmetric, four-dimensional gauge-Yukawa theories with interacting conformal fixed points within a precise framework. The theories investigated in this work are structurally similar to the standard model of particle interactions, but differ by developing perturbative interacting fixed points. We investigate the physical properties of the singlet  and the adjoint composite operators quadratic in the Higgs field, and discover, via a direct computation, that the singlet anomalous dimension is substantially larger than the adjoint one. The numerical bootstrap results are, when possible, compared to our precise findings associated to the four dimensional conformal field theoretical results. To accomplish this, it was necessary to calculate explicitly the crossing symmetry relations for the global symmetry group SU($N$)$\times$SU($N$).
 \\
[.3cm]
{\footnotesize  \it Preprint: CP$^3$-Origins-2014-027 DNRF90 \& DIAS-2014-27}
\end{abstract}
\maketitle

\section{Introduction} 
The discovery of a Higgs particle at the Large Hadron Collider is a major leap forward towards the construction of a more complete theory of nature. If the discovered particle is the standard model Higgs, it is imperative to understand the gauge dynamics of nonsupersymmetric four-dimensional gauge-Yukawa theories. 

Among all possible quantum field theories, the ones developing quantum conformal fixed points have a central role \cite{Wilson:1971bg,Wilson:1971dh}.   
Quantum chromodynamics is a time-honored example \cite{Gross:1973id,Politzer:1973fx}, where the celebrated property of asymptotic freedom comes from a non-interacting ultraviolet fixed point \cite{Gross:1973id,Politzer:1973fx}. One can also imagine the existence of ultraviolet fixed points that \emph{are} interacting, and this scenario is referred to as asymptotic safety \cite{Weinberg:1980gg}.

Recently, phenomenological ultraviolet conformal extensions of the standard model with and without gravity have received much attention  
\cite{Kazakov:2002jd,Gies:2003dp,Morris:2004mg,Fischer:2006fz,
Fischer:2006at,Kazakov:2007su,Zanusso:2009bs,
Gies:2009sv,Daum:2009dn,Vacca:2010mj,Calmet:2010ze,Folkerts:2011jz,
Bazzocchi:2011vr,Gies:2013pma,Antipin:2013exa,Dona:2013qba}. More generally, model building requiring scale invariance both in particle physics and cosmology \cite{Bonanno:2001xi,Meissner:2006zh,Foot:2007ay,Foot:2007iy,Hewett:2007st,Litim:2007iu,Shaposhnikov:2008xi,
Shaposhnikov:2008xb,Shaposhnikov:2009pv,Weinberg:2009wa,
Hooft:2010ac,Gerwick:2011jw,Gerwick:2010kq,
Hindmarsh:2011hx,Hur:2011sv,Dobrich:2012nv,Tavares:2013dga,
Tamarit:2013vda,Abel:2013mya,Antipin:2013bya,Heikinheimo:2013fta,
Gabrielli:2013hma,Holthausen:2013ota,Dorsch:2014qpa,Eichhorn:2014qka} is an active area of research. Furthermore, following Weinberg \cite{Weinberg:1980gg}, even quantum aspects of gravity can be addressed in an asymptotic safety scenario \cite{Litim:2011cp,Litim:2006dx,Niedermaier:2006ns,Niedermaier:2006wt,
Percacci:2007sz,Litim:2008tt,Reuter:2012id}. However, in four dimensions, asymptotic safety has only recently \cite{Litim:2014uca} been guaranteed to occur in calculable nonsupersymmetric gauge-Yukawa theories. Last but not least, perturbative and non-perturbative infrared interacting fixed points are very interesting, both theoretically and phenomenologically \cite{Sannino:2009za,Appelquist:1998xf,Lenaghan:2001sd,Dietrich:2006cm,Ryttov:2007sr,Sannino:2008pz, Sannino:2009aw,Jarvinen:2009fe,Antipin:2009dz,Fukano:2010yv,Ryttov:2010hs,Ryttov:2010iz, Sannino:2010ca,Mojaza:2010cm,Pica:2010mt,Pica:2010xq,Antipin:2011ny,Ryttov:2012ur,Mojaza:2012zd, Foadi:2012ga,Antipin:2012sm,Ryttov:2012nt,Ryttov:2013hka,Antipin:2013qya,Ryttov:2013ura, Molgaard:2014mqa,Shrock:2014qea}. For infrared non-perturbative fixed points in gauge theories, lattice computations are making remarkable progress \cite{
Catterall:2007yx,Catterall:2008qk,DelDebbio:2008tv,DelDebbio:2009fd,
Deuzeman:2009mh,DelDebbio:2010hx,DelDebbio:2010hu,Fodor:2009wk,Fodor:2009ar,Fodor:2009ff, Bursa:2011ru,Hietanen:2008mr,Appelquist:2011dp,Hietanen:2009az,Hietanen:2009zz,Bursa:2009tj, Bursa:2009we,DelDebbio:2013hha}.

In this work, we therefore wish to press forward and investigate explicit conformal properties of nonsupersymmetric gauge-Yukawa theories. We are particularly interested in the properties associated with enforcing crossing symmetry on four-point correlation functions. The microscopic theories investigated here are $SU(N_c)$ gauge theories featuring $N_f$ Dirac fermions transforming according to the fundamental representation of the gauge group, $\ell$ adjoint Weyl fermions, and $N_f^2$ complex scalars, encapsulated in the Higgs matrix $H$. The scalars are coupled to the fermion and gauge sectors via Yukawa interactions. The existence of Banks-Zaks (BZ) \cite{Banks:1981nn} interacting fixed points in such a model has been established in \cite{Antipin:2011aa,Antipin:2012kc,Antipin:2012sm,Antipin:2013pya,Antipin:2013sga}. Furthermore, in  \cite{Litim:2014uca} the reader will find an in depth study of the asymptotic safety scenario and crucial properties which are guaranteed to exist for some of these theories. In this case, the underlying gauge theory is fundamental even in the presence of elementary scalars \cite{Litim:2014uca}.  

Having nonsupersymmetric, interacting, four-dimensional conformal field theories (CFTs) at our disposal, we determine the physical properties of the singlet ${\rm Tr}[H H^{\dagger}]$ and the adjoint ${\rm Tr}[T^a H T^a H^{\dagger}]$ composite operators. Via an explicit computation, we discover that the singlet anomalous dimension is substantially larger than the adjoint one. We then construct the four-point correlations functions in which these operators play an important role, and check  the crossing relations. Furthermore in the Veneziano limit, and at the maximum known order in perturbation theory, we argue that the singlet sector of the theory is nontrivial. We finally compare, when possible, our precise results with the numerical bootstrap constraints \cite{Rattazzi:2008pe,Rychkov:2009ij,Poland:2011ey,ElShowk:2012hu,Nakayama:2014lva}.

The work is organized as follows. In Section \ref{CBS} we briefly review the conformal bootstrap idea and the associated bounds \cite{Rattazzi:2008pe,Rychkov:2009ij,Poland:2011ey}. We then move on to derive the conformal bootstrap sum rules in a CFT with non-Abelian global symmetry $SU(N_f) \times SU(N_f)$ in Section \ref{CBSs}. The four dimensional gauge-Yukawa theories used here are introduced in Section \ref{theory}. In the same section we also argue that the singlet sector decouples from the other operators. In Section \ref{concl}, we offer our conclusions.

\section{Conformal bootstrap review}
\label{CBS}
To set the stage, we provide a short, self-contained introduction to the idea of the conformal bootstrap and highlight its salient properties.   
We consider the set of correlation functions for all local operators of some quantum field theory. For this to constitute a conformal field theory, the set of correlation functions must obey a corresponding set of constrains, and presently, we set out to find it. A CFT consists of its conformal primary operators ${\cal O}_i$\footnote{Primary operators are annihilated by generators of special conformal transformation $[K_{\mu},\mathcal{O}(0)]=0$ where we inserted primary operator at $x=0$ point and $K_{\mu}$ denotes the generator of the special conformal transformation.}, and their
associated conformal dimensions $\Delta_i$ and spins $l_i$. Because of conformality, the normalization is completely arbitrary, and we select a basis for the scalar operators such that the 2-point functions have the form
\beq
\langle {\cal O}_i (x)\, {\cal O}_j(y) \rangle = {\delta_{ij} \over |x-y|^{2\Delta_i}} \ . \label{gen2pf}
\eeq
$\Delta_i$ must satisfy the unitarity constraints \cite{Mack:1975je}:
\beq
\Delta_i &\geq& 1 \qquad \qquad  \quad \ \ (l_i=0)\\
\Delta_i &\geq& l_i+2 \qquad \qquad  (l_i\geq 1)
\eeq
In any CFT, it is possible to express the product of two local operators as a sum over all local operators in the theory which have a finite radius of convergence. This is called the \emph{operator product expansion} (OPE), and we have
\beq
{\cal O}_i(x) \,{\cal O}_j(y) =\sum_k c_{ij}^k(x-y) {\cal M}_k(y) \,
\label{genOPE}
\eeq
where, as mentioned, the sum is over all (primary and non-primary) local operators ${\cal M}_k$ and $c_{ij}^k(x-y)$ are functions of the dimensions and spins (which we denote collectively by the index $k=(\Delta_k,l_k)$) of the operators involved, and of the dynamics of the theory. Using equation \eqref{genOPE} inside correlation functions, we can replace a product (like the LHS) by a sum (like the RHS), as long as there are no other operators at smaller distances from $y$ than $|x-y|$.

The OPE above is quite general, and by also imposing conformal invariance it can be shown \cite{df} that the kinematics of the primary operators uniquely determines the coefficients $c_{ij}^k(x-y)$ belonging to their descendant operators\footnote{The descendant operators are obtained by acting on the primaries with the translation operator i.e. taking derivatives of the primaries.}. Thus, all dynamical information in the OPE is encoded in the coefficients for the primary operators
\beq
{\cal O}_i(x) \, {\cal O}_j(y) = \sum_k C_{ij}^k {\cal O}_k{1\over
|x-y|^{\Delta_i+\Delta_j - \Delta_k}} +{\rm descendants \ contribution},
\eeq
where the new coefficients $C_{ij}^k$ are translation invariant constants. The complete OPE (with both primary and descendant contributions) is then
\beq
{\cal O}_i(x) {\cal O}_j(y) = \sum_k C_{ij}^k \, \ L_k(x-y,\partial_y) \,{\cal O}_k(y) 
{1\over
|x-y|^{\Delta_i+\Delta_j - \Delta_k}}
\eeq 
where $L_k(x-y,\partial_y)$ are differential operators that only depend on the kinematics, that is the dimensions and spins of the primary operators ${\cal O}_k$. They do not depend on the dynamics of the CFT. By using the OPE on the two operators that are closest together, it is now a straightforward matter to reduce an $n$-point function to an infinite sum over $(n-1)$-point functions, which in turn can be reduced to an infinite sum of $(n-2)$-point functions, and so on down to the 2-point functions, which have the simple structure seen in \eqref{gen2pf}. Thus, if we know the conformal dimensions $\Delta_i$, the spins $l_i$ and the 3-point coefficients $C_{ij}^k$ of the primary operators, we know the entire CFT. 

If we have multiple operators, there are several ways of using the OPE to reduce an $n$-point function. However, this obviously cannot change the result, and thus we must insist that regardless of the order in which multiple OPE contractions are used, the end results must be equal. This leads to non-trivial constraints on the possible values of $\Delta_i$ and $C_{ij}^k$ that can make up a consistent CFT. For comprehensive review on both of these constraints see \cite{Poland:2011ey}.

As an instructive example, we consider the 4-point function $\langle {\cal O}_1(x_1){\cal O}_2(x_2){\cal O}_3(x_3){\cal O}_4(x_4) \rangle$. We can evaluate this using the OPE between the operators at $x_1$ and $x_2$ and simultaneously at $x_3$ and $x_4$, or alternatively by performing the OPE between the operators at $x_1$ and $x_4$ and simultaneously at $x_2$ and $x_3$. This corresponds to the s-channel $(12)\to(34)$ and t-channel $(14)\to(23)$ respectively.\footnote{It is also possible to make the contractions in the u-channel $(13)\to(24)$, but this gives no additional constraints.} The contraction in the s-channel yields
\beq
\langle {\cal O}_1(x_1){\cal O}_2(x_2){\cal O}_3(x_3){\cal O}_4(x_4) \rangle
= \sum_k \frac{C_{12}^k C_{34}^k L_k(x_{12},\partial_{x_2})L_k(x_{34},\partial_{x_4})
\langle {\cal O}_k (x_{2})\, {\cal O}_k(x_{4}) \rangle} {|x_{12}|^{\Delta_1+\Delta_2 - \Delta_k}
\ |x_{34}|^{\Delta_3+\Delta_4 - \Delta_k}}\ .
%{1\over |x_2-x_4|^{2\Delta_k}} 
\eeq 
In this expression, only the OPE coefficients $C_{12}^k$ and $C_{34}^k$ depend on the dynamics of the CFT. It is therefore convenient to define the \emph{conformal blocks}
\beq
{\bf G}_k^{12,34}(x_1,x_2,x_3,x_4) \equiv {1\over
|x_{12}|^{\Delta_1+\Delta_2 - \Delta_k}} {1\over
|x_{34}|^{\Delta_3+\Delta_4 - \Delta_k}}L_k(x_{12},\partial_{x_2})L_k(x_{34},\partial_{x_4}) 
\langle {\cal O}_k (x_2)\, {\cal O}_k(x_4) \rangle \  ,%=\frac{L_k(x_{12},\partial_{x_2})L_k(x_{34},\partial_{x_4})}{x_{24}^{2\Delta_k}} 
\label{CB}
\eeq
which contain every contribution from the local operator ${\cal O}_k$ and its many descendants. As mentioned above, these conformal blocks are dependent only on the kinematics of the conformal group, and explicit expressions for them are given in \cite{Dolan:2000ut,Dolan:2003hv}.

The above evaluation was done in the s-channel $(12)\to(34)$, but we could equally well have performed it in the t-channel $(14)\to(23)$. This would have given us a similar, but distinct, expression with 2 and 4 interchanged. Imposing that these two procedures give equal expressions is what yields the non-trivial \emph{conformal bootstrap equation}
\beq
\sum_k C_{12}^k C_{34}^k\,\, {\bf G}_k^{12,34}(x_1,x_2,x_3,x_4) = \sum_k C_{14}^k C_{23}^k \,\, {\bf G}_k^{14,23} (x_1,x_4,x_2,x_3)\ ,
\eeq
which, together with \eqref{CB}, tells us how the dimensions, spins and OPE coefficients must relate to each other in order for the theory in question to be conformal.

In addition, conformal symmetry allows us to further constrain the coordinate dependence of the 4-point function and 
the most general conformally invariant expression is%
\begin{eqnarray}
\langle {\cal O}_1(x_1){\cal O}_2(x_2){\cal O}_3(x_3){\cal O}_4(x_4) \rangle&=&\sum_k C_{12}^k C_{34}^k\,\, {\bf G}_k^{12,34}(x_1,x_2,x_3,x_4) \nonumber \\
&\equiv&\left(  \frac{|x_{24}|}{|x_{14}|}\right)  ^{\Delta_{1}-\Delta_{2}%
}\left(  \frac{|x_{14}|}{|x_{13}|}\right)  ^{\Delta_{3}-\Delta_{4}}%
\frac{g(u,v)}{|x_{12}|^{\Delta_{1}+\Delta_{2}}|x_{34}|^{\Delta_{3}+\Delta_{4}%
}}\ , \label{4pt}%
\end{eqnarray}
where $g(u,v)$ is an arbitrary function\footnote{Note that we absorbed the OPE coefficients $C_{12}^k$ and $C_{34}^k$ into the definition of $g(u,v)$.} of the conformally-invariant
cross-ratios:
\begin{equation}
u=\frac{x_{12}^{2}x_{34}^{2}}{x_{13}^{2}x_{24}^{2}}\ \ ,\quad v=\frac{x_{14}%
^{2}x_{23}^{2}}{x_{13}^{2}x_{24}^{2}} \ \ .
\end{equation}

%Let $\phi\equiv\phi_{d}$ be a Hermitean scalar primary\footnote{The field is
%called primary if it transforms homogeneously under the $4D$ conformal group.}
%operator with the conformal dimension $d$. The operator product expansion (OPE) $\phi\times\phi$ contains, in
%general, infinitely many primary fields of arbitrary high spins and
%dimensions:%
%\beq
%\phi(x)\phi(0)\sim\frac{1}{|x|^{2d}}\left\{  \mathds{1}+\sum_{l=2n}%
%c_{\Delta,l}\left[  \vphantom{\sum}|x|^{-\Delta}K_{l}(x)\cdot O_{\Delta
%,l}(0)+\cdots\right]  \right\}  \,, \quad K_{l}(x)=\frac{x^{\mu_{1}}\cdots
%x^{\mu_{l}}}{|x|^{l}}\,. \label{eq-OPE}%
%\eeq
%Here
%\begin{itemize}
%\item $l=2n$ by Bose symmetry;
%
%\item $\Delta\geq1$ $(\Delta\geq l+2)$\ for $l=0$ ($l\geq2$) by the unitarity
%bounds;
%
%\item The $\cdots$\ stands for contributions of descendants of the primary
%$O_{\Delta,l}$ (i.e.~its derivatives). These contributions are fixed by
%conformal symmetry;
%
%\item The OPE coefficients $c_{\Delta,l}$ are real.
%\end{itemize}
%
In \cite{Rattazzi:2008pe}, the bootstrap equation for the 4-point function of four \emph{identical} scalar operators $\left\langle \phi\phi\phi
\phi\right\rangle $ was considered. 
Starting from the OPE:
\beq
\phi(x)\phi(0)=\frac{1}{x^{2d}}(1+C_{\phi\phi}|x|^\Delta \phi^2(0)+\dots) \ , \quad \ \   d\equiv \Delta_{\phi} \ ,
\label{OPEscalar}
\eeq
and using \eqref{4pt} with all $\Delta_i=d$ equal, we obtain:%
\begin{align}
&  \left\langle \phi(x_{1})\,\phi(x_{2})\,\phi(x_{3})\,\phi(x_{4}%
)\right\rangle =\frac{g(u,v)}{x_{12}^{2d}\,x_{34}^{2d}}\,,\label{eq-4pt}\quad \quad\\
&  g(u,v)
%\equiv x_{12}^{2d}\,x_{34}^{2d}\sum_k C_{12}^k C_{34}^k\,\, {\bf G}_k^{12,34}(x_1,x_2,x_3,x_4) 
=1+\sum p_{k}\,g_{k}(u,v)\,,\quad p_{k}\equiv
(C_{\phi\phi}^k)^{2}\geq0, \label{eq-guv}%
\end{align}
where 
%$u\equiv x_{12}^{2}x_{34}^{2}/(x_{13}^{2}x_{24}^{2}),$ $v\equiv x_{14}^{2}x_{23}%
%^{2}/(x_{13}^{2}x_{24}^{2})$ are the conformal cross-ratios  and 
we explicitly separated the contribution of the identity operator. The
explicit expression for the conformal blocks $g_{k}(u,v)$ reads:%
\begin{align}
&  g_{k}(u,v)=g_{\Delta,l}(u,v)=\frac{(-1)^{l}}{2^{l}}\frac{z\bar{z}}{z-\bar{z}}\left[
\,k_{\Delta+l}(z)k_{\Delta-l-2}(\bar{z})-(z\leftrightarrow\bar{z})\right]
\,,\label{gk-gDl}\\[0.14in]
&  k_{\beta}(x)\equiv x^{\beta/2}{}_{2}F_{1}\left(  \beta
/2,\beta/2,\beta;x\right)  \,,\label{eq-DO}   \quad\qquad u=z\bar{z},\quad v=(1-z)(1-\bar{z})\,.\nonumber
\end{align}
where ${}_{2}F_{1}$ is Gauss's hypergeometric function.

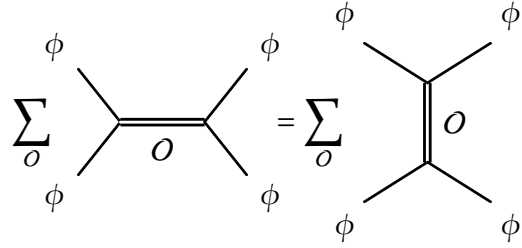
\begin{figure}[ptbh]
\begin{center}
 \begin{fmffile}{bootstrap-diagrams}
\begin{equation*}
\sum_{\mathcal{O}}
\vcenter{\hbox{\begin{fmfgraph*}(80,40)
    \fmfleft{i1,i2}
    \fmfright{o1,o2}
    \fmf{plain}{i1,v1}
    \fmf{plain}{i2,v1}
    \fmf{dbl_plain,label=$\mathcal{O}$}{v1,v2}
    \fmf{plain}{v2,o1}
    \fmf{plain}{v2,o2}
    \fmflabel{$\phi$}{i1}
    \fmflabel{$\phi$}{i2}
    \fmflabel{$\phi$}{o1}
    \fmflabel{$\phi$}{o2}
\end{fmfgraph*}}}
=\sum_{\mathcal{O}}
\vcenter{\hbox{\begin{fmfgraph*}(60,60)
    \fmfleft{i1,i2}
    \fmfright{o1,o2}
    \fmf{plain}{i1,v1}
    \fmf{plain}{i2,v2}
    \fmf{dbl_plain,label=$\mathcal{O}$}{v1,v2}
    \fmf{plain}{v1,o1}
    \fmf{plain}{v2,o2}
    \fmflabel{$\phi$}{i1}
    \fmflabel{$\phi$}{i2}
    \fmflabel{$\phi$}{o1}
    \fmflabel{$\phi$}{o2}
\end{fmfgraph*}}}
\end{equation*}
\end{fmffile}
\caption{Diagrammatic expression of the conformal bootstrap equation. The double line
denotes a conformal block, summing up exchanges of a primary
operator $\mathcal{O}$ and all of its descendants.}%
\label{fig:bootstrap}%
\end{center}
\end{figure}

The 4-point function on the left-hand side of Eq. (\ref{eq-4pt}) is obviously symmetric under the interchange of any two $x_{i}$, and its conformal block decomposition (\ref{eq-guv}) must therefore also respect this symmetry. Invariance with respect to $x_{1}\leftrightarrow
x_{2}$ or $x_{3}\leftrightarrow x_{4}$ implies that only operators of even spin are exchanged.
The non-trivial constraint  comes from the symmetry with respect to
$x_{1}\leftrightarrow x_{3}$ and gives the following condition (see Fig.~\ref{fig:bootstrap} for an illustration)
\begin{equation}
v^{d}g(u,v)=u^{d}g(v,u)\,, \label{eq-boot}%
\end{equation}
which is not automatically satisfied for $g(u,v)$ as given in equation (\ref{eq-guv}).\footnote{
The appearance of the $(u/v)^d$ factor in this relation is due to a nontrivial transformation of the prefactor $1/(x_{12}^{2d}x_{34}^{2d})$ in \eqref{eq-4pt}.}

Following \cite{Rattazzi:2008pe}, it is useful to rewrite (\ref{eq-boot}) by separating the
unit operator contribution, which gives
\begin{equation}
%u^{d}-v^{d}=\sum p_{\Delta,l}\left[  v^{d}g_{\Delta,l}(u,v)-u^{d}g_{\Delta,l}(v,u)\right]  . 
u^{d}-v^{d}=\sum_k p_k\left[  v^{d}g_k(u,v)-u^{d}g_k(v,u)\right]  , 
\label{eq-1sep}%
\end{equation}
where the index $k$ covers the conformal dimension $\Delta$ and the spin $l$, as in \eqref{gk-gDl}.
The LHS of this equation is the imbalance created by the presence of the unit operator in the
OPE. This imbalance has to be compensated by contributions of the other fields on
the RHS.

In practice, it is convenient to normalize (\ref{eq-1sep}) by dividing both
sides by $u^{d}-v^{d}$. The resulting sum rule takes the form:%
\begin{align}
%1 =\sum p_{\Delta,l}F_{d,\Delta,l}\,,\quad \quad F_{d,\Delta,l}   \equiv\frac{v^{d}g_{\Delta,l}(u,v)-u^{d}g_{\Delta,l}(v,u)\,}{u^{d}-v^{d}}. 
1 =\sum_k p_k F_{d,k}\,,\quad \quad F_{d,k}   \equiv\frac{v^{d}g_k(u,v)-u^{d}g_k(v,u)\,}{u^{d}-v^{d}}. 
\label{eq-F}%
\end{align}

For a given spectrum of operator dimensions and spins $\left\{
\Delta,l\right\}  $ the sum rule (\ref{eq-F}) can be viewed as an equation for the
coefficients $p_{\Delta,l}\geq0$. If there are no solutions to this equation, the corresponding CFT would be ruled out.

To achieve a concrete realization of this idea, it is necessary to have a practical recipe to show that
the solution does not exist. For a simple example of such recipe, imagine
that a certain derivative, e.g. $\partial_{x}$, when
applied to every $F_{d,\Delta,l}$ and evaluated at a certain point, is
strictly positive.
Since the same derivative applied to the LHS of (\ref{eq-F}) gives identically
zero, a solution where all coefficients $p_{\Delta,l}$ are non-negative would
clearly be impossible. Using this logic, 
%numerically realizedby linear programming methods, 
a first model-independent bound on the dimension of the operator $\phi^2$ was numerically found in  \cite{Rattazzi:2008pe,Rychkov:2009ij} by using linear programming methods: 
\beq
\Delta \leq \Delta_{min}=2+0.7\sqrt{d-1}+2.1(d-1)+0.43 (d-1)^{3/2} \ ,
\label{OPE1}
\eeq
where $d$ is the conformal dimension of the scalar $\phi$, $d\equiv \Delta_\phi$, and $\Delta$ is the dimension of the operator $\phi^2$, $\Delta\equiv \Delta_{\phi^2}$. 
In \cite{Poland:2011ey} a semidefinite programming algorithm was used and the bound was improved further to the current strongest limit:
\beq
\Delta \leq \Delta_{max}=2+3.006(d-1)+0.16(1-e^{-20(d-1)}) \ .
\label{OPE}
\eeq
There does not seem to be any known 4D unitary CFT saturating this bound.

\section{Conformal bootstrap sum rules in CFT with $SU(N_f)_L\times SU(N_f)_R$ global symmetry}
\label{CBSs}
{ We will construct explicit examples of CFTs stemming from four-dimensional, nonsupersymmetric gauge-Yukawa theories possessing the global, non-Abelian symmetry \mbox{$SU(N_f)_L\times SU(N_f)_R$}. For this reason, we will proceed to generalize the conformal block decomposition to this particular case since it has not, to our knowledge, previously been studied in the literature. Similar analyses have been carried out for the $SO(N)$ and $SU(N)$ cases in \cite{Rattazzi:2010yc}. The relevant gauge singlet complex scalar degrees of freedom are bi-fundamental with respect to the $SU(N_f)_L\times SU(N_f)_R$ global symmetry and can be mathematically represented as:}
\beq
H_i^{{\alpha^*}}={\bf (N_f,N_f^*)}  \ \ \ \  \text{and} \ \ \ \  H^{\beta}_{j^*}={\bf (N_f^*,N_f)} \ \  ,
\eeq 
where all indices $i,j,\alpha,\beta=1,2,\ldots,N_f$. Latin indices are for $SU(N_f)_L$ and Greek indices for $SU(N_f)_R$ respectively. It is convenient to introduce the following matrix notation \beq
H_i^{\alpha^*}=(H)_{i\alpha}   \ \ \ \  \text{and} \ \ \ \  H^{\beta}_{j^*}=(H^{\dagger})_{\beta j} \ \ .
\eeq 
We start with the OPE analysis for the following composite operator: 
%\footnote{\textcolor{red}{In contrast with \eqref{OPEscalar} we do not separate the contribution of the identity operator explicitly.}}
\ea{
\begin{split}
  H_{i\alpha}(x)\times H^\dagger_{\beta j}(0) \sim{}& \frac{1}{|x|^{2d_H}} \bigg\{\delta_{ij}\delta_{\alpha\beta}\left[1+c_S |x|^{\Delta_S}\tr[HH^\dagger](0) \right] \\
  &+c_L|x|^{\Delta_L} { \delta_{ij} }M_{kk\alpha\beta}(0)+c_R|x|^{\Delta_R} {\delta_{\alpha \beta} } M_{ij\gamma \gamma}(0)+c_A|x|^{\Delta_A}M_{ij\alpha\beta}(0)+\cdots\bigg\} \ ,
  \label{basicOPE}
\end{split}
}
where, in the free theory, $M_{ij\alpha\beta}\equiv H_{i\alpha} H^{\dagger}_{\beta j}-\frac{1}{N_f^2}\tr[HH^\dagger]\delta_{ij}\delta_{\alpha\beta}$ and $d_H$ is the conformal dimension of the $H$ field. The group-theoretical content of the OPE above is:
\beq
{\bf (N_f,N_f^*)} \times {\bf (N_f^*,N_f)} = {\bf (1,1)+(1,\text{Adj})+(\text{Adj},1)+(\text{Adj},\text{Adj})} \ . 
\label{group}
\eeq
%We will not need $H\times H$ OPE for our purposes.

The crossing symmetry constraints are derived by equating the (12)$\to$(34) and (14)$\to$(23) s- and t-channel conformal block decompositions of the following 4-point function 
\beq\langle H(x_1)H(x_2)^\dagger H(x_3)H(x_4)^\dagger\rangle=\langle H(x_1)H(x_4)^\dagger H(x_3)H(x_2)^\dagger\rangle \ .
\label{stchannel}
\eeq
There are four basic invariants contained in $[H(x_1)\times H(x_2)^\dagger]\times [H(x_3)\times H(x_4)^\dagger]$.  Using \eqref{group}, we see that the overall singlet terms contributing are:
\ea{
%\begin{split}
&\Bigg\{\bigg[{\bf (1,1)+(1,\text{Adj})+(\text{Adj},1)+(\text{Adj},\text{Adj})} \bigg]\times \bigg[{\bf (1,1)+(1,\text{Adj})+(\text{Adj},1)+(\text{Adj},\text{Adj})} \bigg]\Bigg\}_{\rm singlet} \nonumber\\ 
={}& {\bf G_{S}(1,1) +G_{L}(1,1_{AA})+G_{R}(1_{AA},1)+G_{A}(1_{AA},1_{AA})} 
%\end{split}
}
where ${\bf 1_{AA}}$ means that we have to extract the singlet from the tensor product of the two adjoint representations. In general, each of the four basic invariants $G_{S,L,R,A}$ contain operators of both even and odd spins.
% and the notation $G\equiv \sum p_{\Delta,l} g_{\Delta,l}(u,v)$ means that we have to sum up conformal blocks of all fields with a given symmetry in the s-channel. 

We now derive the constraint stemming from crossing symmetry in terms of these four basic invariants. 
For the s- and t-channel conformal block decompositions we obtain:%
\begin{align}
\begin{split}
\label{s-channelCB}\langle H_1^{\phdagg} H_2^\dagger H_3^{\phdagg} H_4^\dagger\rangle ={}& \frac{1}{x_{12}^{2d_H}x_{34}^{2d_H}}\Bigg\{\left(\raisebox{10pt}{$\xymatrix@=12pt{\ar@{{*}-{o}}[d]& \\ & \ar@{{*}-{o}}[u] }$}\right)^2 G_{S}+\left(  \raisebox{10pt}{$\xymatrix@=12pt{\ar@{{*}-{o}}[r]&
\\\ar@{{o}-{*}}[r]& }$}-\frac{1}{N_f}%
\raisebox{10pt}{$\xymatrix@=12pt{\ar@{{*}-{o}}[d]& \ar@{{o}-{*}}[d]
\\& }$}\right)^2  G_{\text{A}}\\
&+\raisebox{10pt}{$\xymatrix@=12pt{\ar@{{*}-{o}}[d]& \\ & \ar@{{*}-{o}}[u] }$}\left(  \raisebox{10pt}{$\xymatrix@=12pt{\ar@{{*}-{o}}[r]&
\\\ar@{{o}-{*}}[r]& }$}-\frac{1}{N_f}%
\raisebox{10pt}{$\xymatrix@=12pt{\ar@{{*}-{o}}[d]& \ar@{{o}-{*}}[d]
\\& }$}\right)G_L+\left(  \raisebox{10pt}{$\xymatrix@=12pt{\ar@{{*}-{o}}[r]&
\\\ar@{{o}-{*}}[r]& }$}-\frac{1}{N_f}%
\raisebox{10pt}{$\xymatrix@=12pt{\ar@{{*}-{o}}[d]& \ar@{{o}-{*}}[d]
\\& }$}\right)\raisebox{10pt}{$\xymatrix@=12pt{\ar@{{*}-{o}}[d]& \\ & \ar@{{*}-{o}}[u] }$}G_R \Bigg\} \ ,
\end{split}\\
\begin{split}
\langle H_1^{\phdagg} H_4^\dagger H_3^{\phdagg} H_2^\dagger\rangle ={}& \frac{1}{x_{14}^{2d_H}x_{23}^{2d_H}}\Bigg\{  \left(\raisebox{10pt}{$\xymatrix@=12pt{\ar@{{*}-{o}}[r]&
\\\ar@{{o}-{*}}[r]& }$}\right)^2 \widetilde{G}_{S}+\left(
\raisebox{10pt}{$\xymatrix@=12pt{\ar@{{*}-{o}}[d]& \ar@{{o}-{*}}[d]
\\& }$}-\frac{1}{N_f}\raisebox{10pt}{$\xymatrix@=12pt{\ar@{{*}-{o}}[r]&
\\\ar@{{o}-{*}}[r]& }$}\right)^2  \widetilde{G}_{\text{A}}\\
&+\raisebox{10pt}{$\xymatrix@=12pt{\ar@{{*}-{o}}[r]&
\\\ar@{{o}-{*}}[r]& }$}\left(
\raisebox{10pt}{$\xymatrix@=12pt{\ar@{{*}-{o}}[d]& \ar@{{o}-{*}}[d]
\\& }$}-\frac{1}{N_f}\raisebox{10pt}{$\xymatrix@=12pt{\ar@{{*}-{o}}[r]&
\\\ar@{{o}-{*}}[r]& }$}\right)\widetilde{G}_L +\left(
\raisebox{10pt}{$\xymatrix@=12pt{\ar@{{*}-{o}}[d]& \ar@{{o}-{*}}[d]
\\& }$}-\frac{1}{N_f}\raisebox{10pt}{$\xymatrix@=12pt{\ar@{{*}-{o}}[r]&
\\\ar@{{o}-{*}}[r]& }$}\right)\raisebox{10pt}{$\xymatrix@=12pt{\ar@{{*}-{o}}[r]&
\\\ar@{{o}-{*}}[r]& }$}\widetilde{G}_R\Bigg\} \ ,
\end{split}
\end{align}
where $H_i=H(x_i)$, $d_H$ is the quantum physical dimension of the $H$ field, $\widetilde{G}\equiv G_{u\leftrightarrow v}$
%=\sum p_{\Delta,l} g_{\Delta,l}(v,u)$ 
and we used a graphical notation for the tensor contractions. The {\it squaring} of the contractions $(\ldots)^2$ means that we have to perform the same contraction for both $SU(N_f)$ factors. Every line means that the corresponding indices are contracted with the $\delta-$tensor:%
\begin{equation}
\raisebox{12pt}{$\xymatrix@=18pt{\ar@{{*}-{o}}[d]& \\ & \ar@{{*}-{o}}[u] }$}=\delta_{ij}\,\delta_{\alpha\beta}\,,\quad \text{etc.}
%\raisebox{12pt}{$\xymatrix@=18pt{\ar@{{*}-{o}}[r]& \\ \ar@{{o}-{*}}[r]& }$}=\delta_{i\alpha}\,\delta_{j\beta}\,,\text{ etc.}%
\end{equation}
Now, equating the s- and t-channel decompositions and demanding that the coefficients multiplying the corresponding tensors match, we deduce:
\beq
&\left(\raisebox{10pt}{$\xymatrix@=12pt{\ar@{{*}-{o}}[d]& \\ & \ar@{{*}-{o}}[u] }$}\right)^2&\ \ : \quad v^{d_H} \left(G_{S}-\frac{1}{N_f}(G_L+G_R)+\frac{1}{N_f^2}G_{\text{A}}\right)= u^{d_H} \ \widetilde{G}_{\text{A}} \label{single} \ \ , \\ 
&\left(\raisebox{10pt}{$\xymatrix@=12pt{\ar@{{*}-{o}}[r]&
\\\ar@{{o}-{*}}[r]& }$}\right)^2&\ \ : \quad v^{d_H} \ G_{\text{A}} =u^{d_H} \left(\widetilde{G}_{S}-\frac{1}{N_f}(\widetilde{G}_{L}+\widetilde{G}_{R})+\frac{1}{N_f^2}\widetilde{G}_{\text{A}}\right)\label{single1} \ \ ,\\ 
&  \raisebox{10pt}{$\xymatrix@=12pt{\ar@{{*}-{o}}[r]&
\\\ar@{{o}-{*}}[r]& }$}\raisebox{10pt}{$\xymatrix@=12pt{\ar@{{*}-{o}}[d]& \\ & \ar@{{*}-{o}}[u] }$}&\ \ : \quad v^{d_H} \ \left( G_R-\frac{1}{N_f}G_{\text{A}} \right)= u^{d_H} \ \left(\widetilde{G}_{\text{L}}-\frac{1}{N_f}\widetilde{G}_A\right) \ \ ,\label{single3}\\
& \raisebox{10pt}{$\xymatrix@=12pt{\ar@{{*}-{o}}[d]& \\ & \ar@{{*}-{o}}[u] }$} \raisebox{10pt}{$\xymatrix@=12pt{\ar@{{*}-{o}}[r]&
\\\ar@{{o}-{*}}[r]& }$}&\ \ : \quad v^{d_H} \ \left( G_L-\frac{1}{N_f}G_{\text{A}} \right)= u^{d_H} \ \left(\widetilde{G}_{\text{R}}-\frac{1}{N_f}\widetilde{G}_A\right) \label{single4}\ \ ,
 \eeq
which yields four equations with four unknowns. 
%Taking the leading terms at large-$N_f$ we arrive at:
%\beq
%v^{d_H} G_{S}&=& u^{d_H}\widetilde{G}_{A} \quad \quad v^{d_H} G_{A} = u^{d_H}\widetilde{G}_{S} \label{SA}\\
%v^{d_H} G_{L}&=& u^{d_H} \widetilde{G}_R \quad \quad v^{d_H} G_{R}= u^{d_H} \widetilde{G}_L
%\label{split}
 %\eeq 
These equations generalize \eqref{eq-boot} to the theories possessing the non-abelian symmetry $SU(N_f)_L\times SU(N_f)_R$ and can be solved numerically. The 4-point function $\langle H(x_1)H(x_2)^\dagger H(x_3)H(x_4)^\dagger\rangle$ can also be expanded in the (13)$\to$(24) u-channel. For completeness, in Appendix \ref{u-channelsum} we derive the crossing symmetry constraints using this channel while in a follow-up study we will analyze the system \eqref{single}-\eqref{single4}.

%The leading correction to the large-$N_f$ limit appear at the $\mathcal{O}(1/N_f^2)$ order. Indeed, notice that the leading $\mathcal{O}(1/N_f)$ correction to the sum of the Eqs.~(\ref{single}-\ref{single1}) from the $L-R$ conformal blocks vanishes due to the equation \eqref{split}.

\section{A four-dimensional calculable Gauge-Yukawa CFT  }
\label{theory}
 We consider an $SU(N_c)$ gauge theory with $N_f$ fundamental Dirac fermions  $Q = (q, \widetilde{q}^*)$, $\ell$ adjoint Weyl fermions $\lambda$, and a gauge singlet complex scalar $H$ that transforms in the bifundamental representation of the $SU(N_f)_L \times SU(N_f)_R$ global symmetry of the theory. For the benefit of the reader, the field content and the quantum symmetries of the theory with $\ell =1$ are summarized in Table~\ref{FieldContent}.  The hermiticity property of $H$ is defined as $(H_i^{{\alpha^*}})^\dagger \equiv H_{\alpha^*}^{i}$ and 
the matrix $H$ may be decomposed in terms of $2N_f^2$ real scalar fields as follows:
\beq
H_{i}^{\alpha^{\ast}}=H_{i\alpha}= 
\frac{\phi+i\eta}{\sqrt{2N_f}}\delta_{i\alpha}+\sum_{A=1}^{N_f^2-1} (h^A+i\pi^A)T^A_{i \alpha}
\label{matrixH}
\eeq
where $T^A_{i \alpha}$ are the usual generalized Gell-Mann matrices. The fields $H$ and $H^\dagger$ can be contracted to form a singlet
\beq
{\bf (1,1)}=\delta_{ij}\delta_{\alpha\beta} H_{i\alpha} H^{\dagger}_{\beta j}=\tr[HH^\dagger]  \ ,
\eeq
or an adjoint with respect to the right or left handed groups:
\beq
{\bf (\text{\bf Adj},1)}=H_{i\alpha} H^{\dagger}_{\alpha j}=(H H^{\dagger})_{ij} \quad \quad \text{or}\quad \quad {\bf (1,\text{\bf Adj})}=H_{i\alpha} H^{\dagger}_{\beta i}=(H H^{\dagger})_{\alpha\beta} \ ,
\eeq
while {\bf (\text{Adj},\text{Adj})} can be formed as a tensor product  .

The Lagrangian of the theory is
\ea{
\mathcal{L}&= \tr \left[- \frac{1}{2} F^{\mu \nu}F_{\mu \nu} +i \bar{\lambda}  \slashed{D} \lambda+ \overline{Q} i \slashed{D} Q + \partial_\mu H ^\dagger \partial^\mu H + y_H \overline{Q} H Q  \right]  - u_1 (\tr [H H ^\dagger ])^2 -u_2\tr[H H ^\dagger H H^\dagger] . 
 \label{eq:Llsm}
}
Here $\tr$ refers to the trace over both color and flavor indices and ${D_\mu}$ is the usual covariant derivative. At the renormalizable level we have the double trace $(\tr [H H^\dagger])^2$ and the single trace $\tr[H H^\dagger H H^\dagger]$ operators. 

%both of them will contribute to the anomalous dimension of the $\tr[H H^\dagger]$ operator starting already at the lowest one-loop order.
% corresponding to the tensor product of two singlets and the singlet from the tensor product of the two adjoint representations respectively. 

%This is schematically shown for the s-channel on Fig.~\ref{bootstrapAMS} where singlet and adjoint flavor contractions are shown by the arrows.
\begin{table}[t]
\vspace{3mm}
\caption{Field content of the example. The first three fields are Weyl spinors in the ($\frac{1}{2},0$) representation of the Lorentz group. $H$ is a complex scalar and $G_\mu$ is the gauge field. $U(1)_{AF}$ is the extra Anomaly Free symmetry
arising due to the presence of $\lambda$.}%
\vspace{-5mm}
\[ \begin{array}{c|c|c c c c} \hline \hline
{\rm Fields} &\left[ SU(N_c) \right] & SU(N_f)_L &SU(N_f)_R & U(1)_V& U(1)_{AF} \\ \hline 
%\hline 
\lambda & {\rm Adj} & 1 & 1 & 0 & 1 \\
 q &\yf &\overline{\yf }&1&~~\frac{N_f-N_c}{N_c} & - \frac{N_c}{N_f}  \\
\widetilde{q}& \overline{\yf}&1 &  {\yf}& -\frac{N_f-N_c}{N_c}& - \frac{N_c}{N_f}     \\
 %M  & 1 & \yf & \overline{\yf} & 0 & -\frac{N_f-2N_c}{N_f} \\
 \hline
  H & 1 & \yf & \overline{\yf} & 0 & \frac{2N_c}{N_f}\\
  %\wt{\phi} & \overline{\yf} & 1 & \yf & -\frac{N_f-N_c}{N_c} & \frac{N_f-N_c}%{N_f}\\
 % %\phi & \yf & \overline{\yf} & 1 & \frac{N_f-N_c}{N_c} & \frac{N_f-N_c}%{N_f}\\
  G_\mu & \text{Adj} & 1 & 1 & 0 & 0 \\
   \hline \hline \end{array}% 
\]%
\label{FieldContent}%
\vspace{-5mm}
\end{table}
 
Throughout this section we will work with the rescaled couplings which enable a finite Veneziano limit of the theory at fixed $\ell$. That is, we let both $N_c\ , N_f \to \infty$ while keeping $x \equiv N_f/N_c$ fixed. The  appropriately rescaled couplings are
\ea{
a_g = \frac{g^2N_c}{(4 \pi)^2} \ ,~\ a_H = \frac{y_H^2N_c}{(4 \pi)^2}\ ,~ z_1 = \frac{u_1N_f^2}{(4 \pi)^2}\  , ~z_2 = \frac{u_2N_f}{(4 \pi)^2} \ .
}
This model was introduced in \cite{Antipin:2011aa,Antipin:2012sm} to investigate near--conformal dynamics and its impact on the spectrum of the theory. Special attention was paid to the appearance of a dilaton, the Goldstone boson associated with the breaking of conformal symmetry, and its properties. The model was further investigated at higher orders in \cite{Antipin:2012kc}, and the properties related to the {\it a-theorem} were considered in \cite{Antipin:2013pya}.

 \subsection{Beta functions and Weyl consistency conditions}

In order to perform a four-dimensional comparison with the bootstrap bound, we start by providing a calculable CFT at the highest known perturbative order. Following previous studies \cite{Antipin:2013pya,Antipin:2013sga} the beta functions of the theory are
\begin{align}
  \begin{split}
    \label{betaag}\beta_{a_g} ={}& -\frac{2}{3} a_g^2 \left[11-2\ell-2 x+\left(34-16\ell-13 x\right)a_g +\textcolor{red}{3x^2 a_H} +\frac{81x^2}{4}a_g a_H\right.\\
    &\quad\left. -\frac{3x^2(7+6x)}{4}a_H^2+\frac{2857+112x^2-x(1709-257 \ell)-1976 \ell +145\ell^2}{18}a_g^2  \right] \ , 
  \end{split}\\
  \begin{split}
    \label{beta1H}\beta_{a_H} ={}& a_H  \left[ 2(x+1) a_H - \textcolor{red}{6 a_g} +(8 x+5) a_g a_H+\frac{20 (x+\ell)-203}{6} a_g^2\right.\\
    &\quad\left.-\textcolor{blue}{8 x z_2 a_H}-\frac{x(x+12)}{2} a_H^2+\textcolor{brown}{4 z_2^2}\right] \ ,
  \end{split} \\
  \beta_{z_1} ={}& 4 \left(z_1^2+3 z_2^2+4 z_1 z_2 +z_1a_H\right) \ , \\
  \label{beta2Y}\beta_{z_2} ={}& 2 \left(\textcolor{brown}{2 z_2 a_H}+4 z_2^2-\textcolor{blue}{x a_H^2}\right)\ .
\end{align}
Here we have already assumed the Veneziano limit and $\ell$ is the number of $SU(N_c)$ adjoint Weyl fermions of the theory\footnote{In Table~\ref{FieldContent} we assumed $\ell = 1$.}. We used the results of \cite{Machacek:1983tz, Machacek:1983fi, Machacek:1984zw,Luo:2002ti} to determine the beta functions and anomalous dimensions of the gauge-Yukawa theories investigated here. 

The perturbative gauge beta function is considered up to and including the three loop order, the Yukawa to two and the scalar quartic couplings to the first order. This is the proper way of organizing perturbation theory for a multiple coupling theory as shown in 
\cite{Antipin:2013pya,Antipin:2013sga}. In fact this counting can be  mathematically related to the Weyl consistency conditions unveiled in the pioneering work by Osborn \cite{Osborn:1991gm} and demonstrated to be relevant also for the standard model in \cite{Antipin:2013pya,Antipin:2013sga}. These conditions require the different beta functions to be related across different loop orders. Mathematically these conditions read:
\begin{equation}
\frac{\partial (\chi^{jk} \beta_k)}{\partial g_i}=\frac{\partial (\chi^{i m} \beta_m)}{\partial g_j} \ ,\end{equation}
with
\begin{equation}
\chi^{ij}\equiv 
\text{diag}\left( \frac{N_c^2}{128 \pi^2 a_g^2} \ , \frac{N_f^2}{384 \pi^2 a_H}\ , 0\ , \frac{N_f^2}{192 \pi^2}\right)  \ ,
\end{equation}
where $g_i\equiv (a_g, \alpha_H , z_1, z_2)$ refers to the couplings collectively. 
To help the reader identify the related terms, according to the Weyl conditions, across the different couplings, we color--coded them directly in the beta functions. It is clear that these conditions relate the two--loop coefficients in the gauge beta function with one--loop coefficients in the Yukawa beta function (red color) and the two--loop coefficients in the Yukawa beta function with the one--loop coefficients in the quartic beta function (blue and brown colors).  Our perturbative interacting CFTs live at the fixed point (FP) identified by the simultaneous zeros of the previous beta functions, i.e. we need to solve for $\beta_{a_g}=\beta_{a_H}=\beta_{z_1}=\beta_{z_2}=0$. The study of the beta functions above allowing us to establish the existence of perturbative CFTs has been performed in \cite{Antipin:2013pya,Litim:2014uca}. We will investigate  the explicit physical results stemming from the analysis of these beta functions in  \ref{physresults} while in Appendix \ref{bVl} we review, for completeness, the leading finite $N_f$ corrections to the beta functions \cite{Antipin:2012kc}.

\subsection{Higgs anomalous dimensions}

%To the best of our knowledge, the conformal bootstrap bound exploiting the full $SU(N_f)\times SU(N_f)$ global symmetry has not yet been derived. 

The existence of a perturbative CFT permits us to determine the conformal dimensions of the $(1,1)$ singlet $\Delta_S\equiv 2+\gamma_S$ and of the (Adj,Adj) adjoint $\Delta_A\equiv 2+\gamma_A$ composite operators. For the reader's convenience, we recall how these dimensions enters the OPE \eqref{basicOPE}:
\beq
H_{i\alpha}(x)\times H^\dagger_{\beta j}(0) &\sim& \frac{1}{|x|^{2d_H}} \bigg\{ \delta_{ij}\delta_{\alpha\beta}\left[ 1+ c_S |x|^{\Delta_S}\tr[HH^\dagger](0)\right]+c_A|x|^{\Delta_A}M_{ij\alpha\beta}(0)+\cdots\bigg\} \  .
%\phi_{i}\times\phi_{j}  &  =\yng(2)\text{~'s (even spins)}+\yng(1,1)\,\text{'s
%(odd spins)}\,. \label{eq:sun-ope2}%
\label{basicOPE1}
\eeq
To compute these anomalous dimensions, we add to the Lagrangian \eqref{eq:Llsm} two mass terms $m_S^2 \tr[H H^\dagger]$ and $m_A^2 \tr[T^a H T^a H^\dagger]$\footnote{Using the $\mathrm{SU}(N)$ generator identity $T^a_{ij}T^a_{kl}=\frac{1}{2}\delta_{il}\delta_{jk}-\frac{1}{2N}\delta_{ij}\delta_{kl}$, it is easy to see that $\tr[T^a H T^a H^\dagger]=\frac{1}{2}\tr[H]\tr[H^\dagger]-\frac{1}{2N}\tr[HH^\dagger]=\frac{1}{2}M_{ijij}$ and thus how it is related to $M_{ij\alpha\beta}$.} and use \cite{Molgaard:2014hpa} to specialize the formulae given in \cite{Luo:2002ti} to the present case.  

We know from the Weyl consistency conditions \cite{Antipin:2013pya,Antipin:2013sga} that the order to which beta functions are computed in a gauge-Yukawa theory is distinctly non-trivial, and we must therefore also take care to compute the anomalous dimensions of the composite operators, as well as the Higgs field, to the proper order. To find this, we consider that if two of the four external legs on a Feynman digram that contributes to the quartic beta function are joined together, the resulting diagram is a constituent of the anomalous dimension of the composite operators to one higher order in the loop expansion. We therefore conclude that the anomalous dimensions should be computed to two loop order.

Thus, for the Higgs field $H$, we have that the anomalous dimension is
\ea{ 
\gamma_{H}\equiv d_H-1&=a_H+2z_2^2\bigg(1+\frac{1}{N_f^2}\bigg)-\frac{3x a_H^2}{2}+\frac{5 a_g a_H}{2}\bigg(1-\frac{x^2}{N_f^2}\bigg)+2z_1^2\left(\frac{1}{N_f^2}+\frac{1}{N_f^4}\right)+\frac{8z_1z_2}{N_f^2} \ .
%=\frac{1}{2}\gamma_A  \ .
\label{anomH}
}
and for the singlet and the adjoint composite operators:
\begin{subequations}\label{AD}
\ea{
\gamma_A &= \gamma_{\tr[T^{a} H T^{a} H^\dagger]} \equiv \Delta_A-2=2 \gamma_H+\frac{4z_1}{N_f^2}-\frac{8 a_H z_1}{N_f^2}
-4 z_1^2\left(\frac{2}{N_f^2}+\frac{6}{N_f^4}\right)-\frac{32z_1 z_2}{N_f^2}-24\frac{z_2^2}{N_f^2}  \label{gammaTHTH} \\
\gamma_S &= \gamma_{\tr[H H^\dagger]} \equiv \Delta_S-2=\gamma_A +4(z_1+2z_2)-8 a_H (z_1+2 z_2)-24 z_2^2-\frac{16z_1^2}{N_f^2}-\frac{64z_1 z_2}{N_f^2} \ .\label{gammaHH}
}
\end{subequations}
In Appendix \ref{bVl} we show, for completeness, the leading finite $N_f$ corrections to these anomalous dimensions.

Having precisely computed, for the first time, the anomalous dimensions of relevant composite operators in this theory,  it would be interesting to compare them with the bootstrap analysis. Such a comparison is, however, hampered by the fact that the analytic bootstrap conditions, we derived for  $SU(N_f)_L\times SU(N_f)_R$, have not yet been solved numerically, like it is instead the case for  SU($N$) or SO($N$) global symmetries \cite{Poland:2011ey}.  In this initial exploration we will use partial simplifications occurring in the Veneziano limit of the theory to compare our precise results with some of the existing  numerical bounds. 

%In particular, this would shed light on the question of whether the observed behavior for the singlet anomalous dimension holds in the nonperturbative limit.

\subsection{Bootstrap in the Veneziano limit}

Interestingly, in the Veneziano limit, the conformal dimension of the (Adj, Adj) operator factorizes $\Delta_A=2 d_H$ ($\gamma_A=2\gamma_H$), suggesting that, to two-loop order and in the Veneziano limit, we can identify $M_{ij\alpha\beta}(0)$ in \eqref{basicOPE1} with the operator 
\beq M_{ij\alpha\beta}(0) \sim \ :H_{j\beta} H_{\alpha i}^\dagger: (0) \ ,
\label{DTO}
\eeq
where we define the normal-ordered product $:\,:$ of two operators as the non-singular part of the OPE in the limit where the two space-time points are brought together. Because the anomalous dimension of the adjoint is twice that of the $H$ field, this sector of the theory enjoys properties resembling those of a {\it generalized free scalar}  $H(x)$ with conformal dimension $d_H=1+\gamma_H$. Therefore the correlation functions involving the composite adjoint operator are disconnected and can be written as products of 2-point functions. For example, using \eqref{DTO}:
\beq
\langle H_{i\alpha}(x_1)H^\dagger_{\beta j}(x_2) M_{ij\alpha\beta}(y) \rangle = \langle H_{i\alpha}(x_1)H^\dagger_{\alpha i}(y)  \rangle \ \  \langle H_{j\beta}(y)H^\dagger_{\beta  j}(x_2)  \rangle \ ,
\label{factor}    
\eeq
we can compute the 4-point function using the basic 2-point function:
\beq
\langle  H_{i\alpha}(x)  H^\dagger_{\alpha i}(0) \rangle =\frac{1}{|x|^{2d_H}} \  .
\label{2ptGaussian}
\eeq
Moreover, since the 3-point function, defining the OPE coefficient $c_A$ in \eqref{basicOPE1}, is fixed ($\Delta_1 =\Delta_2 = d_H$ and $\Delta_y = \Delta_A =2d_H\ $)
\beq
\langle H_{i\alpha}(x_1)H^\dagger_{\beta j}(x_2) M_{ij\alpha\beta}(y) \rangle = \frac{c_A}{|x_{12}|^{\Delta_1+\Delta_2-\Delta_y}|x_{1y}|^{\Delta_1+\Delta_y-\Delta_2}|x_{2y}|^{\Delta_2+\Delta_y-\Delta_1}}= \frac{c_A}{|x_{1y}|^{2 d_H} |x_{2y}|^{2d_H}}
\label{forappendix}
\eeq 
by comparing with \eqref{factor} and using \eqref{2ptGaussian} we see that $c_A=1$. 

The factorization property of the (Adj, Adj) operators allows us to compute $G_A$ and $\widetilde{G}_A$ to this order in perturbation theory and in the Veneziano limit. Indeed, to compute $G_A$, for example, we start with the general expressions \eqref{4pt} (with all $\Delta_i=d_H$) and using notation of \eqref{s-channelCB} write:
\begin{align}
\begin{split}
\label{explicit-4pt1}
\langle H_{i\alpha}(x_1)H^\dagger_{\beta j}(x_2)H_{k\delta}(x_3)H^\dagger_{\sigma m}(x_4) \rangle = 
\Bigg[ \left(\raisebox{10pt}{$\xymatrix@=12pt{\ar@{{*}-{o}}[r]&
\\\ar@{{o}-{*}}[r]& }$}\right)^2 G_A \Bigg] \cdot \frac{1}{x_{12}^{2d_H}x_{34}^{2d_H}}+ \cdots
\end{split}
\end{align}
where we showed explicitly only the contributions from the conformal block $G_A$. As indicated by index contractions, we have to consider the correlator with external indices ($i=m$, $j=k$) and ($\alpha=\sigma$, $\beta=\delta$). Using the factorization property of (Adj,Adj), we calculate the $G_A$ contribution as follows \cite{Heemskerk:2009pn}\footnote{Here we assume a complete factorisation in the (Adj, Adj) channel although we have shown that it holds only for the leading operator. Therefore we have the generalized free Gaussian theory with OPE \cite{Heemskerk:2009pn}: 
\beq H(x_1)\times H^\dagger (x_2) =\frac{1}{x_{12}^{2d_H}}+\sum_{n,l} \frac{c_{n,l}^{A}}{x_{12}^{n+l}} O_{n,l}^{A}
\eeq
where only double-trace operators $\mathcal{O}_{n,l}^{A}=(\mathcal{O}^{A}\overleftrightarrow{\partial_{\mu_1}}...\overleftrightarrow{\partial_{\mu_l}}(\overleftrightarrow{\partial_{\nu}}\overleftrightarrow{\partial^{\nu}})^n \mathcal{O}^{A} $ - traces)  contribute.
%$\mathcal{O}_{n,l}^{A}=\mathcal{O}^{A}\overleftrightarrow{\partial_{\mu_1}}...\overleftrightarrow{\partial_{\mu_l}}(\overleftrightarrow{\partial_{\nu}}\overleftrightarrow{\partial^{\nu}})^n \mathcal{O}^{A} $ - traces 
%\\ \\
Here $l$ is the spin of the operator and $\Delta_{n,l}=2 d_H + 2n+l+\mathcal{O}(1/N_f^2)$.  
%Contributions of the higher-trace operators are absent and the OPE coefficients $c_{n,l}^{A}$ can be found from the OPE above. 
The leading operator $O_{0,0}^{A}=Tr[T^a H T^a H^\dagger]=\frac{1}{2}Tr[H] Tr[H^\dagger]+\mathcal{O}(1/N_f)$ has dimension $2d_H$. }
\beq
[\langle H_{i\alpha}(x_1)H^\dagger_{\beta j}(x_2)H_{k\delta}(x_3)H^\dagger_{\sigma m}(x_4) \rangle]_{G_A}= 
% \langle H_{i\alpha}(x_1)H^\dagger_{\sigma m}(x_4)  \rangle \ \  \langle H_{k\delta}(x_3)H^\dagger_{\beta  j}(x_2)  \rangle = 
\frac{1}{x_{14}^{2d_H}x_{23}^{2d_H}}
\label{Gablock}
\eeq
and therefore by comparing with \eqref{explicit-4pt1} we deduce that :
\beq
G_A= \bigg(\frac{u}{v}\bigg)^{d_H}=\frac{x_{12}^{2d_H}x_{34}^{2d_H}}{x_{14}^{2d_H}x_{23}^{2d_H}} \  .
\eeq
Similarly, for $\widetilde{G}_A$ we obtain $\widetilde{G}_A=(v/u)^{d_H}$. In terms of Feynman diagrams, factorization implies that the conformal block $G_A$ contributes only to the disconnected diagrams to this order in perturbation theory and in the Veneziano limit. These disconnected contributions provide the leading-$N_f$ dependence of the correlators which is known as {\it large-$N_f$ factorization} \cite{Makeenko:1999hq}. In fact, using  the standard 't Hooft counting, it is easy to show that the disconnected contribution to our 4-point function in \eqref{Gablock} appear at the $\mathcal{O}(1)$ while the fully connected contributions appear at the $\mathcal{O}(1/N_f^2)$. 
 
{To take advantage of the large-$N_f$ factorization, following \cite{Heemskerk:2009pn}, we will be solving our bootstrap conditions \eqref{single}-\eqref{single4} in the $1/N_f$ expansion:
\beq
G_{S,A}&\equiv& \sum_{\Delta,l} p_{\Delta,l}^{S,A}g_{\Delta,l}^{S,A}(u,v)=G_{S,A}^{disc} + \frac{G_{S,A}^{conn}}{N_f^2}+\cdots \label{Pol1}\\
G_{L,R}&\equiv& \sum_{\Delta,l} p_{\Delta,l}^{L,R}g_{\Delta,l}^{L,R}(u,v)= \frac{G_{L,R}}{N_f}+\cdots \label{Pol2}
\eeq
}
where we formally divided the connected (conn) and disconnected (disc) contributions \cite{Heemskerk:2009pn} to the conformal blocks $G_S$ and $G_A$. 
We also used the fact that $G_L$ and $G_R$ appear at the order  $\mathcal{O}(1/N_f)$ because they are disconnected with respect to just one of the two $SU(N_f)$ factors.

\begin{figure}
\begin{fmffile}{disconneted-diagrams}
\begin{align*}
  \left(\vcenter{\hbox{\rule[-15pt]{0pt}{90pt}\rule{10pt}{0pt}\begin{fmfgraph*}(80,60)
    \fmfleft{i1,i2}
    \fmfright{o1,o2}
    \fmflabel{$H$}{i1}
    \fmflabel{$H^\dagger$}{i2}
    \fmflabel{$H^\dagger$}{o1}
    \fmflabel{$H$}{o2}
    \fmf{dashes}{i1,v1}
    \fmf{dashes}{i2,v1}
    \fmf{dashes}{v1,o1}
    \fmf{dashes}{v1,o2}
    \fmfv{decor.shape=circle,decor.filled=30,decor.size=.4h}{v1}
\end{fmfgraph*}}}\rule{10pt}{0pt}\right)_{\text{disconnected}} &= 
\hspace{10pt}\vcenter{\hbox{\begin{fmfgraph*}(80,60)
    \fmfleft{i1,i2}
    \fmfright{o1,o2}
    \fmflabel{$H$}{i1}
    \fmflabel{$H^\dagger$}{i2}
    \fmflabel{$H^\dagger$}{o1}
    \fmflabel{$H$}{o2}
    \fmf{dashes}{i1,v1}
    \fmf{dashes}{i2,v2}
    \fmf{dashes}{v1,o1}
    \fmf{dashes}{v2,o2}
    \fmfv{decor.shape=circle,decor.filled=30,decor.size=.4h}{v1}
    \fmfv{decor.shape=circle,decor.filled=30,decor.size=.4h}{v2}
\end{fmfgraph*}}}\hspace{15pt}
+
\hspace{15pt}\vcenter{\hbox{\begin{fmfgraph*}(60,60)
    \fmfleft{i1,i2}
    \fmfright{o1,o2}
    \fmflabel{$H$}{i1}
    \fmflabel{$H^\dagger$}{i2}
    \fmflabel{$H^\dagger$}{o1}
    \fmflabel{$H$}{o2}
    \fmf{dashes}{i1,v1}
    \fmf{dashes}{i2,v1}
    \fmf{dashes}{v2,o1}
    \fmf{dashes}{v2,o2}
    \fmfv{decor.shape=circle,decor.filled=30,decor.size=.4h}{v1}
    \fmfv{decor.shape=circle,decor.filled=30,decor.size=.4h}{v2}
\end{fmfgraph*}}}
\end{align*}
\end{fmffile}
\caption{Disconnected contributions to the 4-point function.}
\label{fact2}
\end{figure}
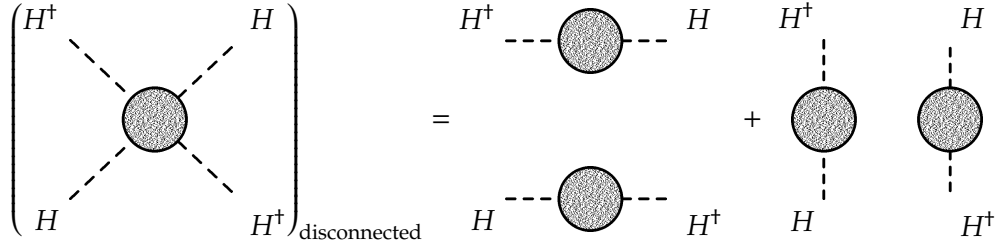

At the leading $\mathcal{O}(1)$ in the large-$N_f$ expansion, from the bootstrap equations \eqref{single}-\eqref{single1} we have:
\beq
\mathcal{O}(1) : \qquad u^{d_H}\widetilde{G}_{A}^{disc}&=&v^{d_H} G_{S}^{disc}  \ , \quad {\rm with} \quad \widetilde{G}_A^{disc}=\bigg(\frac{v}{u}\bigg)^{d_H} \ , \label{s-ch}\\
\mathcal{O}(1) : \qquad v^{d_H} G_{A}^{disc}&=& u^{d_H}\widetilde{G}_{S}^{disc} \ ,  \quad {\rm with} \quad G_A^{disc}=\bigg(\frac{u}{v}\bigg)^{d_H} \label{t-ch}\ ,
\eeq
which means that $G_S^{disc}=\widetilde{G}_S^{disc}=1$. These equations exemplify crossing symmetry constraints for the disconnected contributions to the 4-point function 
represented schematically in Fig.~\ref{fact2}. 
%The explicit expressions for the OPE coefficients $(p_{\Delta,l}^0)^{S,A}$ solving the bootstrap constraints at this leading $\mathcal{O}(1)$ order can be found in \cite{Heemskerk:2009pn}.

The relevant bootstrap conditions at $\mathcal{O}(1/N_f^2)$ are derived by combining the conditions \eqref{single}-\eqref{single4} with the expansion \eqref{Pol1}-\eqref{Pol2}, $G_A^{disc}=(u/v)^{d_H}$, $\widetilde{G}_A^{disc}=(v/u)^{d_H}$ and $G_S^{disc}=\widetilde{G}_S^{disc}=1$.  By matching the $1/N_f^2$ terms we have:
\beq
&\mathcal{O}(1/N_f^2) :&  \qquad v^{d_H} \left(G_{S}^{conn}-(G_L+G_R)\right)+u^{d_H}=u^{d_H} \widetilde{G}_A^{conn}  \ ,\label{eqone}\\
&\mathcal{O}(1/N_f^2) :&  \qquad u^{d_H} \left(\widetilde{G}^{conn}_{S}-(\widetilde{G}_{L}+\widetilde{G}_{R})\right)+v^{d_H}=v^{d_H} G_A^{conn} \ ,\label{eqtwo}\\
&\mathcal{O}(1/N_f^2) :&  \qquad \left[ v^{d_H}(G_L+G_R) -u^{d_H}(\widetilde{G}_L+\widetilde{G}_R) \right] = 2( u^{d_H}-v^{d_H}) \label{eqthree}\ .
\eeq
The last equation is obtained by subtracting \eqref{single4} from \eqref{single3}.
Using \eqref{eqthree} in the equation obtained by subtracting \eqref{eqtwo} from \eqref{eqone} we arrive at:
\beq
v^{d_H} G_{S}^{conn}- u^{d_H}\widetilde{G}^{conn}_{S}=u^{d_H}(1+\widetilde{G}_A^{conn})-v^{d_H}(1+G_A^{conn}) \label{finalbootAMS}\ .
\eeq
 {Let us now consider the contributions to the conformal blocks $G_{S,A}^{conn}$ and $\widetilde{G}_{S,A}^{conn}$ more carefully}. We will be using the work of \cite{Heemskerk:2009pn} where the 4-point function of the singlet operators $\mathcal{O}(x)$ was considered. In this case, the lowest dimensional operator, aside from the unit operator, in the $\mathcal{O}\times \mathcal{O}$ OPE is the double trace operator $\mathcal{O}^2$ whose dimension $2 d[\mathcal{O}]+\mathcal{O}(1/N_f^2)$ factorises at the lowest order in $1/N_f$ expansion. 

In our model, from \eqref{AD}, the dimensions of the adjoint operator satisfy the same factorization property and therefore the analysis of \cite{Heemskerk:2009pn} applies. There it was shown that the conformal block $G_{A}^{conn}$ (and $\widetilde{G}_{A}^{conn}$) receives the contributions from the sum of two terms: 
\begin{itemize}
\item $\mathcal{O}(1/N_f^2)$ correction to the OPE coefficients $p_{\Delta,l}^{A}$. We will denote this contribution by $(G_{A}^{conn})^{OPE}$ 
\item $\mathcal{O}(1/N_f^2)$ corrections to the anomalous dimension $\Delta_{A}$ which enter the functions $g_{\Delta,l}^{A}(u,v)$. We will denote this contribution by 
$(G_{A}^{conn})^{AD}$ \ .
\end{itemize}
The connected contribution to the conformal block can thus be expanded \cite{Heemskerk:2009pn}: 
\beq
G_{A}^{conn}=(G_{A}^{conn})^{OPE}+(G_{A}^{conn})^{AD} \ \ . \qquad 
%\text{and} \qquad \widetilde{G}_{A}^{conn}=(\widetilde{G}_{A}^{conn})^{OPE}+(\widetilde{G}_{A}^{conn})^{AD} \ .
\label{notation}
\eeq
Furthermore, in our model the anomalous dimension for the singlet operator from \eqref{AD} equals the anomalous dimension of the adjoint plus an additional \emph{non-factoriziable} contribution $4(z_1+2z_2)-8 a_H (z_1+2 z_2)-24 z_2^2$ and two additional $\mathcal{O}(1/N_f^2)$ terms $-16z_1^2/N_f^2-64z_1 z_2/N_f^2$ not present in the anomalous dimensions for the adjoint \eqref{AD}. 
%These $\mathcal{O}(1/N_f^2)$ terms are related to terms of the same order in the quartic beta functions. This can be easily seen by considering Eqs. \eqref{fullbetaz1}-\eqref{fullbetaz2} in Appendix \ref{bVl}. 
%Since our beta functions in \eqref{betaag}-\eqref{beta2Y} are defined in the Veneziano limit, they are not included there. 
%We will include these corrections later in Section \ref{beyondVL} where we add the $\mathcal{O}(1/N_f^2)$ corrections to our beta functions. 
The non-factorizable contribution is not present in the analysis of \cite{Heemskerk:2009pn} and it will be taken into account. 
%In summary, the singlet anomalous dimension is given by
%\beq
%\gamma_S=\gamma_A+4(z_1+2z_2)-8 a_H (z_1+2 z_2)-24 z_2^2 \ .
%\eeq 

Based on the above discussion it seems reasonable, but should still be proven, that $G_{A}^{conn}$ matches the factorazible part of $G_{S}^{conn}$. Assuming that this is true, the contributions due to the conformal blocks ($(G_{S,A}^{conn})^{AD}$,$(\widetilde{G}_{S,A}^{conn})^{AD}$) and ($(G_{S,A}^{conn})^{OPE}$,$(\widetilde{G}_{S,A}^{conn})^{OPE}$) cancel out in \eqref{finalbootAMS}:
\beq
&v&^{d_H} ((G_{S}^{conn})^{AD}+(G_{S}^{conn})^{OPE})- u^{d_H}((\widetilde{G}^{conn}_{S})^{AD}+\widetilde{G}^{conn}_{S})^{OPE})= \nonumber\\
&u&^{d_H}((\widetilde{G}_A^{conn})^{AD}+\widetilde{G}^{conn}_{A})^{OPE})-v^{d_H}((G_A^{conn})^{AD}+\widetilde{G}^{conn}_{A})^{OPE}) \label{finalbootAMS1}\ .
\eeq
%We determine explicitly $(G_{S,A}^{conn})^{AD}$ and $(\widetilde{G}_{S,A}^{conn})^{AD}$ in appendix \ref{appB}. 
 
%Finally, the remaining 
The non-factorizable contribution to the singlet anomalous dimension quantifies the departure from the Gaussian limit and stems from an additional part of the singlet conformal block $(G_{S}^{conn})^{\rm non-fact}$. This part will not be balanced by an appropriate term associated with the adjoint composite operator in \eqref{finalbootAMS}. This leads to a suggestive bootstrap equation for the non-factorizable part of the singlet
\beq
v^{d_H} (G_{S}^{conn})^{\rm non-fact}- u^{d_H}(\widetilde{G}^{conn}_{S})^{\rm non-fact}=u^{d_H}-v^{d_H} \label{finalnonfact}\ ,
\eeq 
which has precisely the form of \eqref{eq-1sep}, and we can even expand the conformal block in functions of the kinematics of the CFT $(G_S^{conn})^{\rm non-fact} = \sum p^{S,nf}_{\Delta,l} g^{S,nf}_{\Delta,l}(u,v)$. 

Just as in \eqref{eq-1sep}, the right-hand side of \eqref{finalnonfact} is the contribution from the Gaussian part of the theory which is balanced by the left-hand side. Therefore it would be tempting to interpret this result as a bound for $\Delta_S$, similar to the bound on the lowest dimensional operators of the theory coming from Eq. \eqref{eq-1sep}, though holding only to the next-to-leading order in the couplings and in the Veneziano limit. This would mean that for low values of the couplings and high values of $N_f$, $\Delta_S$ should obey the bound given by Eq. \eqref{OPE} with $d=d_H$. 

However, there are several caveats to this suggestive statement that require further investigation. The most pressing is that our expression holds only for a part of $G_S$, and it is not clear how (or even if) a consistency equation on such a part would translate into a bound on $\Delta_S$. 
%Any bound would furthermore be limited to a narrow region of validity in perturbation theory and infinite $N_f$. 
Another concern is that, as pointed out in \cite{Heemskerk:2009pn}, unitarity only implies positivity of $p_{\Delta,l}$ to leading order in $1/N_f$, and if the expansion parameters $p^{S,nf}_{\Delta,l}$ are allowed to take either sign, the large $N_f$ analogue of the proof provided in \cite{Rattazzi:2008pe} would be affected. Finally, the function $G_S$ contains an implicit sum over even and odd spins, while the bound of
\eqref{OPE} is obtained for real scalars, where only even spins enter the crossing symmetry
constraint.

The bound of \cite{Poland:2011ey} applies directly to the adjoint composite operator, the lowest dimensional operator,  without any caveats and as we shall see in the specific examples provided below, the bound is well satisfied. As for the singlet composite operator, a direct comparison with the bound \eqref{OPE} quantifies the extent to which the caveats described above are under theoretical control .  We stress again that a proper comparison requires a dedicated numerical bootstrap analysis for this theory.

In the examples below, we will also see that the anomalous dimension of the singlet composite operator can be substantially larger than that of the adjoint composite operator. Given that the bound on the singlet is unknown, this is a welcome feature which has been long sought after for nonperturbative models of near conformal dynamics used to describe composite Higgs scenarios, see \cite{Sannino:2009za} for a recent review.

\subsection{Physical results}
\label{physresults}
 
Now we review the salient points behind the existence of  perturbative interacting CFTs \cite{Antipin:2013pya,Litim:2014uca} and then determine the physical dimensions of the composite operators at the FPs of the theory.  

The two--loop gauge beta function has a perturbative Banks-Zaks  FP if the one--loop coefficient $b_0$ of the gauge beta function is small and the signs of the one-loop $b_0$ and two-loop $b_1$ coefficients are opposite. Therefore, our first task is to find a region in the parameter space of the model where the BZ FP exists. Solving  \eqref{beta1H} to one-loop and substituting into \eqref{betaag} we obtain
\begin{equation}
\label{tuning}
b_0= \frac{2}{3}\big(11-2(\ell+x)\big) \ ,  \qquad  b_1=\frac{2}{3}\big((34-16\ell-13x)+\frac{9x^2}{x+1}\big) \ .
\end{equation}
From the asymptotic freedom (AF) boundary condition $b_0=0$, we obtain \mbox{$x=(11-2\ell)/2$}. After substituting this value of $x$ into $b_1$
\begin{equation}
\label{tuning1}
b_{1AF}=-\frac{25}{2}-\ell-\frac{3(11-2\ell)^2}{4\ell-26}\ ,
\end{equation}
we observe that for the unphysical value $\ell^*\approx 0.37$ the coefficient $b_{1AF}$ vanishes. For $\ell =1 $ we have that $b_{1AF}$ is negative and for $\ell=0$ it is positive. Therefore in the first case we have an infrared BZ FP, and in the second we have an ultraviolet BZ FP. Note also \cite{Antipin:2013pya} that in the absence of the Yukawa interactions the coefficient $b_{1AF}$ in \eqref{tuning1} is always negative and therefore the physical BZ FP can lead only to an infrared FP. 

We are now ready to present our results for the $\ell=0$ and $\ell=1$ cases. 
{Our strategy is the following}
\begin{itemize}
\item For a given FP in all the couplings $(a_g^*, a_H^*, z_1^*, z_2^*)$ at a given value of $x\equiv N_f/N_c$, we determine the anomalous dimensions for the composite operators $\gamma_{S}$ and $\gamma_A$. We also determine the associated anomalous dimension of the scalar field $\gamma_H$. These results were obtained by means of the equations  \eqref{anomH} and \eqref{AD}.
\item We then insert $d = d_H=1+\gamma_H$ in the right hand-side of \eqref{OPE1} and \eqref{OPE} to determine $\Delta_{max}-2$. Finally, we compare the result with $\gamma_A$, which turns out to be the operator for which the bootstrap bound applies in all cases under consideration, and display $\gamma_S$, which is a more interesting quantity for phenomenology. \end{itemize}

%Because of the factorization property for the adjoint composite operator, i.e. $\gamma_A=2\gamma_H$, it is guaranteed that the physical dimension of the (Adj,Adj) operator satisfies the bound \eqref{OPE} \cite{Rychkov:2009ij}. The situation, as we shall see, is much more intriguing for the scalar operator. 

%Furthermore, because the more recent numerically determined functional form, given in \eqref{OPE} \cite{Poland:2011ey}, has a different dependence of $\gamma_H$ in the very perturbative region than the previous bound \eqref{OPE1} \cite{Rychkov:2009ij}, we also display this for a comparison.
 
\subsubsection{\texorpdfstring{The $\ell=0$ case}{The $l=0$ case}}
\begin{figure}[t!]
\subfloat[Fixed point structure of the model with $\ell=0$. The boundary of asymptotic freedom is on the left-hand edge of the plot at $x=5.5$, the FP value of $a_g$ is the solid red line, $a_H$ is the dotted black, $z_1$ is the dot-dashed green, and $z_2$ is the dashed blue.]{\includegraphics[width=.49\textwidth]{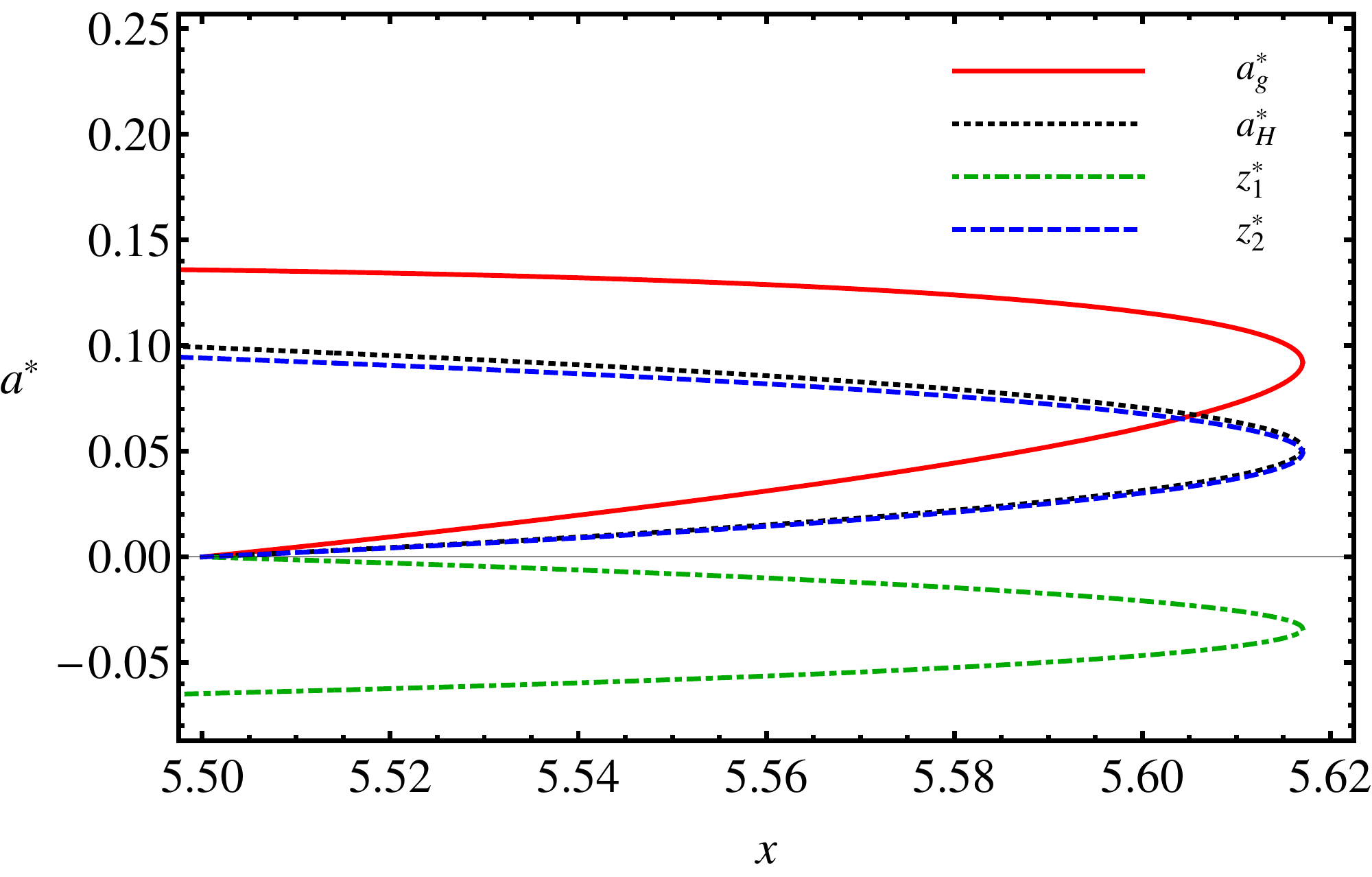}\label{FP0a}}
  \hfill
\subfloat[Comparison with the conformal bootstrap bounds. 
The solid line is the bound \eqref{OPE1}, the dot-dashed black is the revised bound \eqref{OPE}, the dashed gray line is $\gamma_{\tr[HH^\dagger]}$ \eqref{gammaHH}, and the dotted light gray line is $\gamma_{\tr[T^aHT^aH^\dagger]}$ \eqref{gammaTHTH}.]{\includegraphics[width=.49\textwidth]{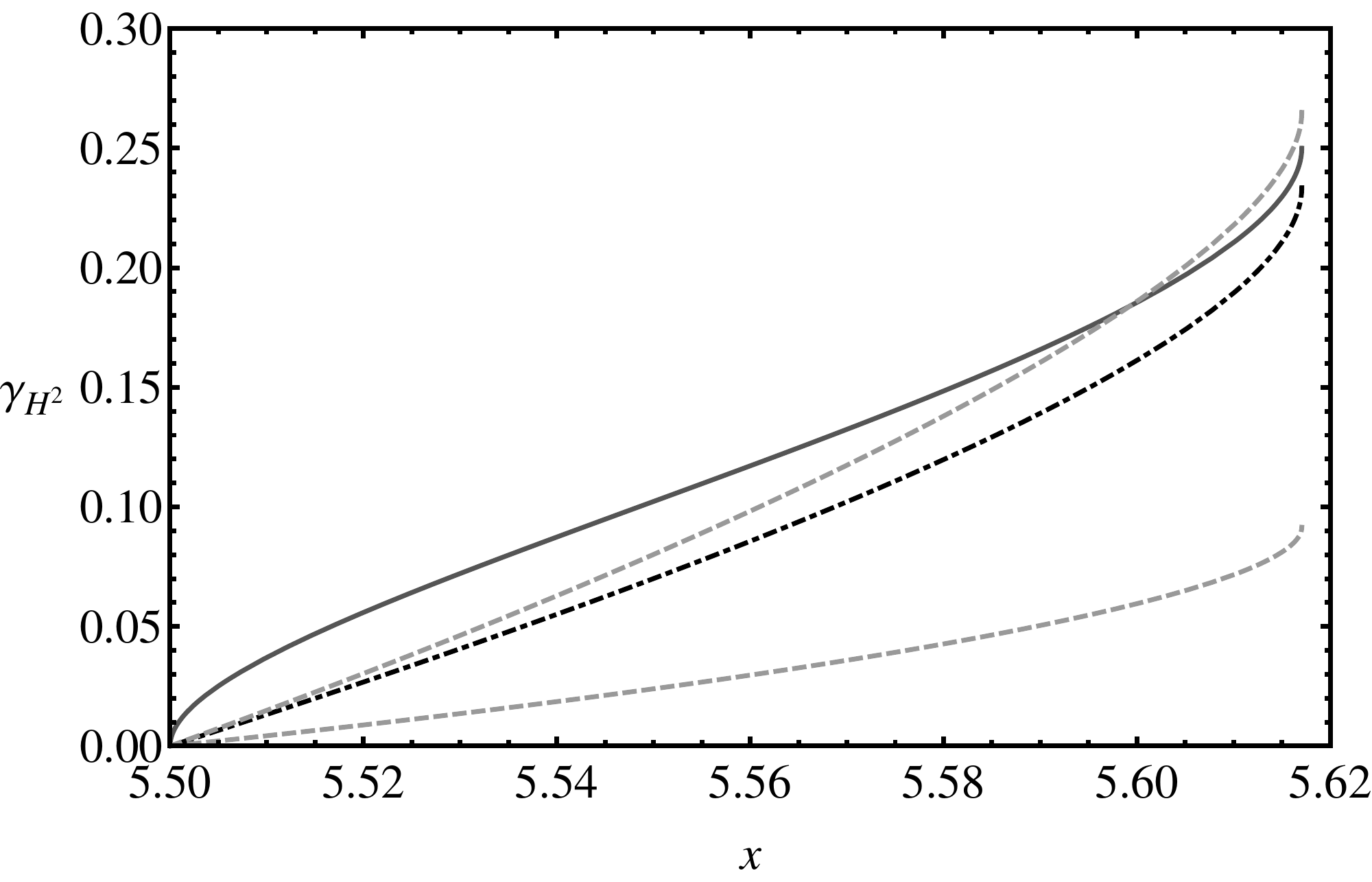}\label{FP0b}}
\caption{FP structure and comparison with the bootstrap bound for the model with $\ell=0$.}
\label{FP0}
\end{figure}

The asymptotic freedom boundary, where the first coefficient of the gauge beta function vanishes, $b_0=0$, occurs at $x_{AF}=(11-2\ell)/2=5.5$. Increasing $x >x_{AF}$ slightly results in the appearance of an ultraviolet BZ FP, see Fig.~\ref{FP0a}. An in depth analysis of the FP structure and its theoretical and phenomenological consequences for the asymptotic safety scenario has just appeared in \cite{Litim:2014uca}. 

When the three--loop gauge beta function is considered, an infrared FP emergers along with the ultraviolet BZ FP in the range $x\le x^*\approx 5.617$. At $x^*$ the ultraviolet BZ FP and the infrared FP collide and both fixed points disappear. Perturbation theory is, of course, valid only for values of $(x-x_{AF})/x_{AF} \ll 1$. As shown in \cite{Litim:2014uca} perturbation theory is valid for $(x-x_{AF})/x_{AF} < 0.1$.

The comparison with the bootstrap bound is shown in Fig.~\ref{FP0b}. We first note that as expected, the bound is clearly respected by the anomalous dimension of the adjoint operator. More interestingly, we discover that the anomalous dimension of the singlet composite operator $\gamma_S$ is substantially larger than the anomalous dimension of the adjoint operator. If this also holds in the non-perturbative regime, this has important and welcome implications for model building. 

\subsubsection{\texorpdfstring{The $\ell=1$ case}{The $l=1$ case}}
\begin{figure}[h!]
\subfloat[Fixed point structure of the model with $\ell=1$. The FP value of $a_g$ is the solid red line, $a_H$ is the dotted black, $z_1$ is the dot-dashed green, and $z_2$ is the dashed blue. ]{\includegraphics[width=.49\textwidth]{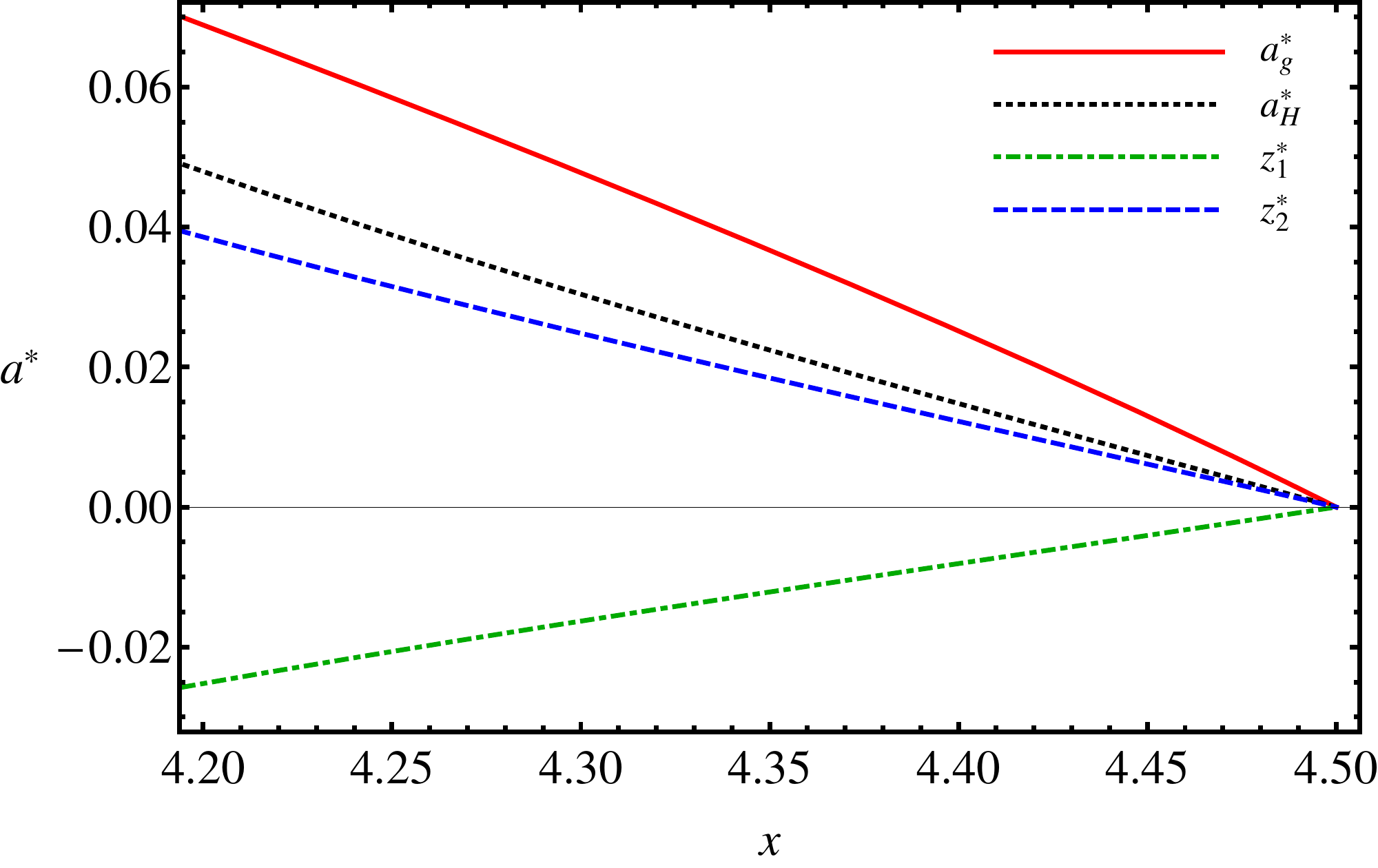}\label{FP1a}} \hfill
 \subfloat[Comparison of the conformal bootstrap bound.
 The solid gray line is the bound \eqref{OPE1}, the dot-dashed black is the revised bound \eqref{OPE}, the dashed gray line is $\gamma_{\tr[HH^\dagger]}$ \eqref{gammaHH}, and the dotted light gray line is $\gamma_{\tr[T^aHT^aH^\dagger]}$ \eqref{gammaTHTH}.]{\includegraphics[width=.49\textwidth]{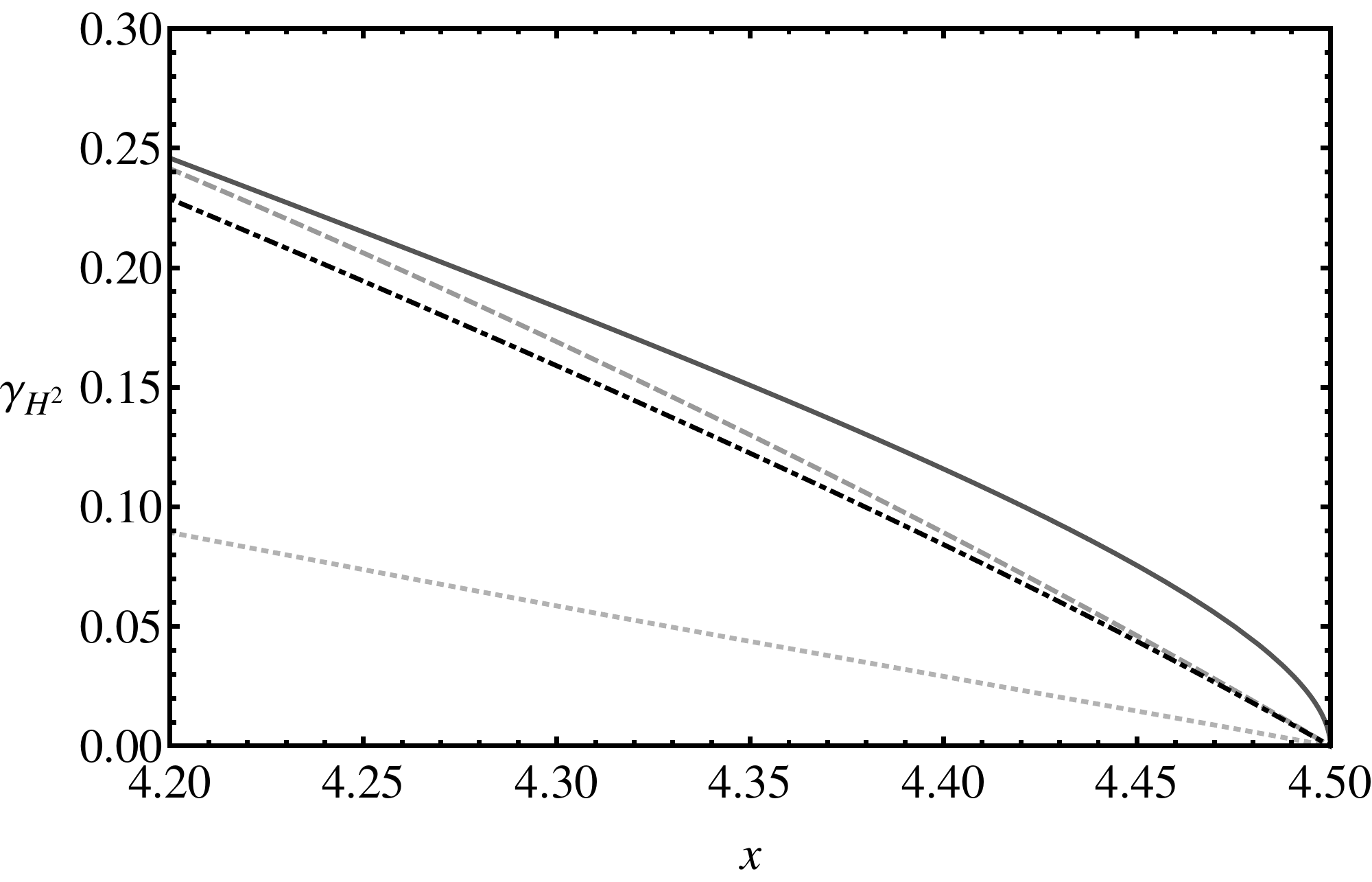}\label{FP1b}}
\caption{FP structure and comparison with the bootstrap bound for the model with $\ell=1$.}
\label{FP1}
\end{figure}

When the model is expanded to include adjoint fermions, the infrared BZ FP originates just below the asymptotic freedom boundary $x_{AF}=(11-2\ell)/2=4.5$ as shown in Fig.~\ref{FP1a}. 
 The comparison of the composite operator anomalous dimensions with the two numerical bootstrap bounds is shown in Fig.~\ref{FP1b}. As explained above, $\gamma_A$ is consistently below the bound in the perturbative regime. 
 {As for the $\ell=0$ case we determine the relevant quantity $\gamma_S$  and show that it is, also in this case, substantially larger than the adjoint operator.  }
\section{Conclusions}
\label{concl}
We provided a systematic investigation of interesting properties of relevant composite operators stemming from gauge-Yukawa theories developing conformal fixed points in four dimensions. These theories are structurally similar to the standard model of particle interactions and have already been employed for interesting model building \cite{Abel:2013mya}. Having at our disposal explicit examples of nonsupersymmetric interacting four-dimensional CFTs, we investigated the critical exponents (anomalous dimensions at fixed points) associated to singlet ${\rm Tr}[H H^{\dagger}]$ and the adjoint ${\rm Tr}[T^a H T^a H^{\dagger}]$ operators. 

We showed that in the Veneziano limit, and at the maximum known order in perturbation theory, the adjoint composite operator is Gaussian and automatically obeys the bootstrap bounds on the anomalous dimension.
%the connected non-trivial scattering contribution to the singlet sector decouples from the other operators and therefore we were able to compare its anomalous dimension with the bootstrap constraints derived for a CFT without global symmetries. 
We also discovered that the  singlet composite operator anomalous dimension at the interacting FP is substantially larger than the one for the  adjoint composite operator. This is an interesting observation for phenomenologically driven questions regarding the possibility of large anomalous dimensions for singlet operators needed, for example, in theories of composite Higgs dynamics \cite{Sannino:2009za}. It would be interesting to analyze more generally the full bootstrap equations for these patterns of chiral symmetry.
 
Our results demonstrate the relevance of constructing conformal nonsupersymmetric four dimensional gauge-Yukawa theories that can be used for demonstrating the existence of four dimensional asymptotically safe theories \cite{Litim:2014uca}, for interesting model building \cite{Abel:2013mya}, probing the {\it a-theorem}  \cite{Antipin:2013pya}, but also to either accurately test numerical solutions of the bootstrap constraints or {determine novel anomalous dimensions of relevant composite operators}.  Following the pioneering work of Seiberg \cite{Seiberg:1994pq} it would be interesting to explore whether the weakly coupled four dimensional gauge-Yukawa theories investigated here have strongly coupled duals \cite{Sannino:2009qc,Sannino:2010fh}.

  \acknowledgments
  
 We thank V. Rychkov for constructive discussions and insights. We also thank M. Redi for valuable discussions.
The CP$^3$-Origins centre is partially funded by the Danish National Research Foundation, grant number DNRF90. The work of OA is supported by the MIUR-FIRB grant RBFR12H1MW.

\newpage
\appendix

\section{ The u-channel sum rule \label{u-channelsum}}

To extract the full information on the 4-point function, we also need to consider the (13)$\to$(24) $u$-channel OPE which reads:
\begin{align}
{\bf H \times H}= {\bf (N_f, N_f^*)\times (N_f, N_f^*)  }
%\sim \yng(2)\text{~'s (even spins)}+\yng(1,1)\,\text{'s
%(odd spins)}\,. 
={\bf (2S,2S^*)^{+}+(2A,2A^*)^{+}+(2S,2A^*)^{-}+(2A,2S^*)^{-}}
\label{eq:sun-ope2}
\end{align}
where $2S$ and $2A$ stand for the two-index symmetric and antisymmetric tensors respectively. Due to the permutation symmetry of the ${\bf H H}$ state, the tensors ${\bf (2S,2S^*)^{+}}$ and ${\bf (2A,2A^*)^{+}}$ contain only even spins while the tensors ${\bf (2S,2A^*)^{-}}$ and ${\bf (2A,2S^*)^{-}}$ contain only odd spins. 

As discussed in \cite{Rattazzi:2010yc}, the additional crossing symmetry constraints can be derived by equating the (12)$\to$(34) and (14)$\to$(23) s- and t-channel conformal block decompositions of the "transposed" ($H(x_3)\rightarrow H(x_3)^\dagger$ and $H(x_4)^\dagger \rightarrow H(x_4)$) 4-point function $\langle H(x_1)H(x_2)^\dagger H(x_3)^\dagger H(x_4)\rangle$: 
\beq\langle H(x_1)H(x_2)^\dagger H(x_3)^\dagger H(x_4)\rangle=\langle H(x_1)H(x_4) H(x_2)^\dagger H(x_3)^\dagger \rangle \ .
\label{stchannel}
\eeq
Now in the t-channel we have the same OPE as we would have in the u-channel of the original 4-point function and in the s-channel we have the same OPE up to the transposition of the fields at spacetime points $x_3$ and $x_4$. This transposition is taken into account by reversing the signs of the odd-spins contributions and permuting the flavor indices accordingly.

There are four basic invariants contained in the t-channel of the "transposed" 4-point function  $[H(x_1)\times H(x_4)]\times [H(x_2)^\dagger\times H(x_3)^\dagger]$.  Using \eqref{eq:sun-ope2}, we see that the overall singlet terms contributing are:
\begin{eqnarray}
&\Bigg[\bigg[ (2S,2S^*)+(2A,2A^*)+(2S,2A^*)+(2A,2S^*)\bigg]\times \bigg[(2S^*,2S)+(2A^*,2A)+(2S^*,2A)+(2A^*,2S)\bigg]\Bigg]_{\rm singlet} \nonumber \\ 
&= {\bf G_{2S,2S} +G_{2A,2A}+G_{2A,2S}+G_{2S,2A}} 
\end{eqnarray}
where ${\bf 1_{2S}}$ and ${\bf 1_{2A}}$ means that we have to extract the singlet from the tensor product of the corresponding two-index representations.

%There are four basic invariants contained in $[H(x_1)\times H(x_2)]\times [H(x_3^\dagger)\times H(x_4)^\dagger]$
%\begin{eqnarray}
%&\Bigg[\bigg[ (2S,2S^*)+(2A,2A^*)+(2S,2A^*)+(2A,2S^*)\bigg]\times \bigg[(2S^*,2S)+(2A^*,2A)+(2S^*,2A)+(2A^*,2S)\bigg]\Bigg]_{\rm singlet} =  \nonumber \\ 
%& {\bf G_{2S,2S} +G_{2A,2A}+G_{2A,2S}+G_{2S,2A}} 
%\end{eqnarray}
%

From the $s=t$ channel crossing symmetry constrain of the "transposed" 4-point function we obtain:
%\begin{equation}
%u^{-d}\left\{  1+G_{S}+G_{\text{Adj}}\right\}  =v^{-d}\left\{1+  G_S +\widetilde{G}_{\text{Adj}}\right\}  , \label{eq:SUN-1}%
%\end{equation}
%
%In addition, equating $s=u$ channels we obtain additional equations (in Veneziano limit):
\begin{align}
\begin{split}
\langle H_1^{\phdagg} H_2^\dagger H_3^{\dagger} H_4^{\phdagg}\rangle &={} \frac{1}{x_{12}^{2d_H}x_{34}^{2d_H}} \Bigg\{
\left(\raisebox{12pt}{$\xymatrix@=18pt{\ar@{{*}-{o}}[d]& \\ & \ar@{{o}-{*}}[u] }$}\right)^2(G_{S}^{+}-G_{S}^{-})+\left(  \raisebox{12pt}{$\xymatrix@=18pt{\ar@{{*}-{o}}[rd]&
\\\ar@{{o}-{*}}[ru]& }$}-\frac{1}{N_f} \raisebox{12pt}{$\xymatrix@=18pt{\ar@{{*}-{o}}[d]& \\ & \ar@{{o}-{*}}[u] }$}\right)^2
(G_{\text{A}}^{+}-G_{\text{A}}^{-})  \\
&+\raisebox{12pt}{$\xymatrix@=18pt{\ar@{{*}-{o}}[d]& \\ & \ar@{{o}-{*}}[u] }$} \left(\raisebox{12pt}{$\xymatrix@=18pt{\ar@{{*}-{o}}[rd]&
\\\ar@{{o}-{*}}[ru]& }$}-\frac{1}{N_f}\raisebox{12pt}{$\xymatrix@=18pt{\ar@{{*}-{o}}[d]& \\ & \ar@{{o}-{*}}[u] }$}\right) (G_L^{+}-G_L^{-})+ \left(\raisebox{12pt}{$\xymatrix@=18pt{\ar@{{*}-{o}}[rd]&
\\\ar@{{o}-{*}}[ru]& }$}-\frac{1}{N_f}\raisebox{12pt}{$\xymatrix@=18pt{\ar@{{*}-{o}}[d]& \\ & \ar@{{o}-{*}}[u] }$}\right) \raisebox{12pt}{$\xymatrix@=18pt{\ar@{{*}-{o}}[d]& \\ & \ar@{{o}-{*}}[u] }$} (G_R^{+}-G_R^{-})\Bigg\} \nonumber
\end{split}\\
\begin{split}
\langle H_1^{\phdagg} H_4 H_2^{\dagger}   H_3^\dagger & \rangle ={} \frac{1}{x_{14}^{2d_H}x_{23}^{2d_H}}\Bigg\{  \left(  \raisebox{12pt}{$\xymatrix@=18pt{\ar@{{*}-{o}}[d]&
\ar@{{*}-{o}}[d] \\&
}$}+\raisebox{12pt}{$\xymatrix@=18pt{\ar@{{*}-{o}}[rd]&
\\\ar@{{o}-{*}}[ru]& }$}\right)^2  \widetilde{G}^{+}_{2S,2S}+\left(
\raisebox{12pt}{$\xymatrix@=18pt{\ar@{{*}-{o}}[d]& \ar@{{*}-{o}}[d]
\\& }$}-\raisebox{12pt}{$\xymatrix@=18pt{\ar@{{*}-{o}}[rd]&
\\\ar@{{o}-{*}}[ru]& }$}\right)^2  \widetilde{G}^{+}_{2A,2A}\\
+&\left(   \raisebox{12pt}{$\xymatrix@=18pt{\ar@{{*}-{o}}[d]&
\ar@{{*}-{o}}[d] \\&
}$}+\raisebox{12pt}{$\xymatrix@=18pt{\ar@{{*}-{o}}[rd]&
\\\ar@{{o}-{*}}[ru]& }$}\right)\left(  \raisebox{12pt}{$\xymatrix@=18pt{\ar@{{*}-{o}}[d]&
\ar@{{*}-{o}}[d] \\&
}$}-\raisebox{12pt}{$\xymatrix@=18pt{\ar@{{*}-{o}}[rd]&
\\\ar@{{o}-{*}}[ru]& }$}\right)\widetilde{G}^{-}_{2S,2A}+\left(  \raisebox{12pt}{$\xymatrix@=18pt{\ar@{{*}-{o}}[d]&
\ar@{{*}-{o}}[d] \\&
}$}-\raisebox{12pt}{$\xymatrix@=18pt{\ar@{{*}-{o}}[rd]&
\\\ar@{{o}-{*}}[ru]& }$}\right)\left(  \raisebox{12pt}{$\xymatrix@=18pt{\ar@{{*}-{o}}[d]&
\ar@{{*}-{o}}[d] \\&
}$}+\raisebox{12pt}{$\xymatrix@=18pt{\ar@{{*}-{o}}[rd]&
\\\ar@{{o}-{*}}[ru]& }$}\right)\widetilde{G}^{-}_{2A,2S}\Bigg\} \nonumber
\end{split}
\end{align}
The s-channel decomposition is obtained from the previous case by transposing
the index structure \textit{and} flipping the sign of the odd-spin
contributions. The t-channel decomposition is obtained by using the second OPE
(\ref{eq:sun-ope2}). The index structure is fixed by (anti)symmetry of the
exchanged fields.
%, while the signs are determined by demanding positive
%contributions for $i=\bar{\imath}\neq j=\bar{\jmath}$ (which makes the
%configuration reflection-positive in the t-channel).
Now, equating the s- and t-channel decompositions and demanding that the coefficients multiplying the corresponding tensors match we deduce:
\beq
 &\left(\raisebox{10pt}{$\xymatrix@=12pt{\ar@{{*}-{o}}[d]& \\ & \ar@{{*}-{o}}[u] }$}\right)^2\ \ : v^{d_H} \left(G_{S}^{+}-G_{S}^{-}+\frac{1}{N_f^2}\big(G_{\text{A}}^{+}-G_{\text{A}}^{-}\big)-\frac{1}{N_f}\big(G_L^{+}-G_L^{-}+G_R^{+}-G_R^{-}\big)\right)= \nonumber \\
& u^{d_H}\big(\widetilde{G}^{+}_{2S,2S}+\widetilde{G}^{+}_{2A,2A}+\widetilde{G}^{-}_{2S,2A}+\widetilde{G}^{-}_{2A,2S}\big) \label{single5} \\ 
&\left(\raisebox{12pt}{$\xymatrix@=18pt{\ar@{{*}-{o}}[rd]&
\\\ar@{{o}-{*}}[ru]& }$} \right)^2 \ \ : v^{d_H} \ \big(G_{\text{A}}^{+}-G_{\text{A}}^{-}\big) =u^{d_H}\big(\widetilde{G}^{+}_{2S,2S}+\widetilde{G}^{+}_{2A,2A}-\widetilde{G}^{-}_{2S,2A}-\widetilde{G}^{-}_{2A,2S}\big)\label{single1a}\nonumber \\ 
& \raisebox{12pt}{$\xymatrix@=18pt{\ar@{{*}-{o}}[rd]&
\\\ar@{{o}-{*}}[ru]& }$}\raisebox{12pt}{$\xymatrix@=18pt{\ar@{{*}-{o}}[d]&
\ar@{{*}-{o}}[d] \\&
}$} \ \ : v^{d_H} \ \bigg(G_R^{+}-G_R^{-}-\frac{1}{N_f}\big(G_{\text{A}}^{+}-G_{\text{A}}^{-}\big)\bigg)  =u^{d_H}\big(\widetilde{G}^{+}_{2S,2S}-\widetilde{G}^{+}_{2A,2A}+\widetilde{G}^{-}_{2S,2A}-\widetilde{G}^{-}_{2A,2S}\big)\nonumber \\
& \raisebox{12pt}{$\xymatrix@=18pt{\ar@{{*}-{o}}[d]&
\ar@{{*}-{o}}[d] \\&
}$}  \raisebox{12pt}{$\xymatrix@=18pt{\ar@{{*}-{o}}[rd]&
\\\ar@{{o}-{*}}[ru]& }$}  \ \ : v^{d_H}\ \bigg(G_L^{+}-G_L^{-}-\frac{1}{N_f}\big(G_{\text{A}}^{+}-G_{\text{A}}^{-}\big)\bigg)  =u^{d_H}\big(\widetilde{G}^{+}_{2S,2S}-\widetilde{G}^{+}_{2A,2A}-\widetilde{G}^{-}_{2S,2A}+\widetilde{G}^{-}_{2A,2S}\big)\nonumber 
%& u^{-d_H} \  G_L= v^{-d_H}(\widetilde{G}_{2S,2S}-\widetilde{G}_{2A,2A}-\widetilde{G}_{2S,2A}+\widetilde{G}_{2A,2S}) \\
%&G_R= v^{-d_H}(\widetilde{G}_{2S,2S}-\widetilde{G}_{2A,2A}+\widetilde{G}_{2S,2A}-\widetilde{G}_{2A,2S}) \ \ ,
 \eeq
Working to lowest order in $1/N_f$ and concentrating on the even spins conformal blocks $G^{+}_{2S,2S}$ and $G^{+}_{2A,2A}$ we have:
\beq
v^{d_H}(G_{S}^{+}+G_{A}^{+}) &=& 2 u^{d_H}(\widetilde{G}^{+}_{2S,2S}+\widetilde{G}^{+}_{2A,2A})
 \eeq
% \xxx{Can we prove that $G_{S}^{-}=-G_{A}^{-}$?}

\section{Complete beta functions and anomalous dimensions beyond the Veneziano limit}

We provide here the full beta functions and anomalous dimensions of the gauge-Yukawa system. The conventions are the ones given in the main text. 

For the beta functions we have:
\ea{
  \begin{split}
    \beta_{a_g}={}&-\frac{2}{3} a_g^2 \Bigg[11-2 x-2 \ell +a_g \left(\frac{3 x^3}{N_f^2}-13 x-16 \ell +34\right)+3 x^2 a_H\\
    &+a_g a_H \left(\frac{81 x^2}{4}-\frac{9 x^4}{4 N_f^2}\right)+a_g^2 \Bigg\{\frac{3x^5}{4 N_f^4}-\frac{11 x^3 (2 x+2 \ell -17)}{12 N_f^2}\\
    &+\frac{1}{18} \left(112 x^2+x (257 \ell -1709)+145 \ell ^2-1976 \ell +2857\right)\Bigg\}-\frac{3}{4} (6x+7) x^2 a_H^2\Bigg]
  \end{split}\\
  \begin{split}
    \beta_{a_H}={}&2 a_H \Bigg[a_g \left(\frac{3 x^2}{N_f^2}-3\right)+(x+1) a_H+a_g^2 \left(-\frac{3 x^4}{4 N_f^4}-\frac{x^2 (5x+5 \ell -53)}{3 N_f^2}+\frac{1}{12} (20 x+20 \ell -203)\right)\\
    &+a_g a_H \left(-\frac{(8 x+5) x^2}{2 N_f^2}+4 x+\frac{5}{2}\right)+a_H^2 \left(\frac{2 x^2}{N_f^2}-\frac{1}{4} x (x+12)\right)\\
    &-\frac{8 x z_1 a_H}{N_f^2}+z_2 a_H\left(-\frac{4 x}{N_f^2}-4 x\right)+z_1^2 \left(\frac{2}{N_f^2}+\frac{2}{N_f^4}\right)+\frac{8 z_1
   z_2}{N_f^2}+z_2^2 \left(\frac{2}{N_f^2}+2\right)\Bigg]
  \end{split}\\
  \beta_{z_1}={}&4 z_1 a_H+z_1^2 \left(\frac{16}{N_f^2}+4\right)+16 z_2 z_1+12 z_2^2 \label{fullbetaz1}\\
  \beta_{z_2}={}&-2 x a_H^2+4 z_2 a_H+\frac{24 z_1 z_2}{N_f^2}+8 z_2^2 \label{fullbetaz2}
}
%The anomalous dimensions at one loop order are:
%\ea{
%  \gamma_H &= a_H
% + a_g a_H \left(\frac{5}{2}-\frac{5 x^2}{2 N_f^2}\right)-\frac{3}{2} x a_H^2+\frac{2 z_1^2 \left(N_f^2+1\right)}{N_f^4}+\frac{2 z_2^2\left(N_f^2+1\right)}{N_f^2}+\frac{8 z_1 z_2}{N_f^2}
%\ ,\\
%  \gamma_S = \gamma_{\tr[HH^\dagger]} &= 2 a_H+z_1 \left(\frac{4}{N_f^2}+4\right)+8 z_2
%+a_g a_H \left(5-\frac{5 x^2}{N_f^2}\right)+z_1 a_H \left(-\frac{8}{N_f^2}-8\right)-3 x a_H^2-16 z_2 a_H+z_1^2 \left(-\frac{20}{N_f^2}-\frac{20}{N_f^4}\right)+z_2^2\left(-\frac{20}{N_f^2}-20\right)-\frac{80 z_1 z_2}{N_f^2}
%\ ,\\
%  \gamma_A = \gamma_{\tr[T^aHT^aH^\dagger]} &= 2 a_H+\frac{4 z_1}{N_f^2}
% + a_g a_H \left(5-\frac{5 x^2}{N_f^2}\right)-\frac{8 z_1 a_H}{N_f^2}-3 x a_H^2+z_1^2 \left(-\frac{4}{N_f^2}-\frac{20}{N_f^4}\right)+z_2^2\left(4-\frac{20}{N_f^2}\right)-\frac{16 z_1 z_2}{N_f^2}
%\ .
%}
And for the anomalous dimensions:
\label{bVl}
\ea{
 \gamma_{H}={}& a_H+2z_2^2\bigg(1+\frac{1}{N_f^2}\bigg)-\frac{3x a_H^2}{2}+\frac{5 a_g a_H}{2}\bigg(1-\frac{x^2}{N_f^2}\bigg)+2z_1^2\left(\frac{1}{N_f^2}+\frac{1}{N_f^4}\right)+\frac{8z_1z_2}{N_f^2}\\ \displaybreak
\begin{split}
  \gamma_S= \gamma_{\tr[H H^\dagger]}={}& 2 a_H +4z_1\bigg(1+\frac{1}{N_f^2}\bigg)+8z_2-3 a_H^2 x+5a_g a_H \left(1-\frac{x^2}{N_f^2}\right)-8a_H z_1 \left(1+\frac{1}{N_f^2}\right)\\
  &-20 z_1^2\left(\frac{1}{N_f^2}+\frac{1}{N_f^4}\right)-16 a_H z_2-\frac{80}{N_f^2}z_1 z_2-20 z_2^2\left(1+\frac{1}{N_f^2}\right)
\end{split}\\
 \gamma_A=\gamma_{\tr[T^a H T^a H^\dagger]}={}& 2 a_H+\frac{4z_1}{N_f^2}-3 a_H^2 x+5 a_g a_H \left(1-\frac{x^2}{N_f^2}\right)-\frac{8 a_H z_1}{N_f^2}\nonumber \\
  &-4 z_1^2 \left(\frac{1}{N_f^2}+\frac{5}{N_f^4}\right)-\frac{16
   z_1 z_2}{N_f^2}+4 z_2^2 \left(1-\frac{5}{N_f^2}\right) \ .
}
Remarkably all the leading $1/N_f$ corrections emerge only at the order $1/N_f^2$ order. 

\end{document}